\documentclass[aps,prb,twocolumn,floatfix,footinbib,showpacs,superscriptaddress]{revtex4}
\usepackage[english]{babel}
\usepackage[utf8]{inputenc}
\usepackage[dvips]{graphicx}
\usepackage{fancyhdr}
\usepackage[active]{srcltx}
\usepackage{setspace}
\usepackage{float}
\usepackage{amsmath}
\usepackage{amsfonts}
\usepackage[]{natbib}
\usepackage{array}
\usepackage{color}
\begin{document}

\title{Realistic multiband {\bf k.p} approach from {\it ab initio} and spin-orbit coupling effects of InAs and InP in wurtzite phase}

\author{Paulo E. Faria Junior}
\email{fariajunior.pe@gmail.com}
\affiliation{S\~ao Carlos Institute of Physics, University of S\~ao Paulo, 13566-590 S\~ao Carlos, S\~ao Paulo, Brazil}

\author{Tiago Campos}
\affiliation{S\~ao Carlos Institute of Physics, University of S\~ao Paulo, 13566-590 S\~ao Carlos, S\~ao Paulo, Brazil}
\affiliation{Institute for Theoretical Physics, University of Regensburg, 93040 Regensburg, Germany}

\author{Carlos M. O. Bastos}
\affiliation{S\~ao Carlos Institute of Physics, University of S\~ao Paulo, 13566-590 S\~ao Carlos, S\~ao Paulo, Brazil}

\author{Martin Gmitra}
\affiliation{Institute for Theoretical Physics, University of Regensburg, 93040 Regensburg, Germany}

\author{Jaroslav Fabian}
\affiliation{Institute for Theoretical Physics, University of Regensburg, 93040 Regensburg, Germany}

\author{Guilherme M. Sipahi}
\affiliation{S\~ao Carlos Institute of Physics, University of S\~ao Paulo, 13566-590 S\~ao Carlos, S\~ao Paulo, Brazil}
\affiliation{Department of Physics, State University of New York at Buffalo, Buffalo, New York 14260, USA}

%===============================================================================

\begin{abstract}

Semiconductor nanowires based on non-nitride III-V compounds can be synthesized 
under certain growth conditions to favor the appearance of wurtzite crystal phase. 
Despite the reports in literature of {\it ab initio} band structures for these 
wurtzite compounds, we still lack effective multiband models and parameter sets 
that can be simply used to investigate physical properties of such systems, for 
instance, under quantum confinement effects. In order to address this deficiency, 
in this study we calculate the {\it ab initio} band structure of bulk InAs and InP 
in wurtzite phase and develop an 8$\times$8 {\bf k.p} Hamiltonian to describe the 
energy bands around $\Gamma$ point. We show that our {\bf k.p} model is robust and 
can be fitted to describe the important features of the {\it ab initio} band structure. 
The correct description of the spin splitting effects that arise due to the lack 
of inversion symmetry in wurtzite crystals, is obtained with the $k$-dependent 
spin-orbit term in the Hamiltonian, often neglected in the literature. All the 
energy bands display a Rashba-like spin texture for the in-plane spin expectation 
value. We also provide the density of states and the carrier density as functions 
of the Fermi energy. Alternatively, we show an analytical description of the conduction 
band, valid close to $\Gamma$ point. The same fitting procedure is applied to the 
6$\times$6 valence band Hamiltonian. However, we find that the most reliable approach 
is the 8$\times$8 {\bf k.p} Hamiltonian for both compounds. The {\bf k.p} Hamiltonians 
and parameter sets that we develop in this paper provide a reliable theoretical 
framework that can be easily applied to investigate electronic, transport, optical, 
and spin properties of InAs- and InP-based nanostructures.

\end{abstract}

\pacs{71.15.Mb, 71.20.-b, 71.20.Mq, 71.70.Ej}

% 71.15.Mb: DFT condensed matter
% 71.20.-b: Band structure
% 71.20.Mq: Semiconductors, elemental, band structure of
% 71.70.Ej: Spin-orbit coupling in condensed matter

\maketitle

%===============================================================================

\section{Introduction}
\label{sec:intro}

In the past decade, the development of low dimensional III-V semiconductor nanostructures 
has witnessed great advances.\cite{Li2006:MatT} For instance, one interesting feature 
that was observed in the synthesis of III-V nanowires is the appearance of wurtzite 
(WZ) crystal phase, instead of the usual zinc-blende (ZB).\cite{Caroff2009:NatNano} 
This created new possibilities of III-V compounds with WZ structure besides the 
well established nitride based materials. Moreover, controlling the growth conditions, 
e.~g., temperature and III/V ratio, it is possible to achieve not only single crystal 
phase nanowires\cite{Vu2013:Nanotech,Pan2014:NL} but also to mix ZB and WZ regions 
with sharp interfaces in the same nanostructure, which is known as polytypism.\cite{Dick2010:NL,Bolinsson2011:Nanotech,Lehmann2013:NL} 
It has been demonstrated that mixed phases greatly affect the physical properties, 
for example, of the light polarization,\cite{Hoang2010:NL,Gadret2010:PRB,FariaJunior2014:JAP} 
electron transport,\cite{Thelander2011:NL,Konar2015:NL} and photoconductivity.\cite{Li2015:APL}

Among these new III-V compounds with WZ structure, InAs and InP are particularly important. 
InAs WZ has a large spin-orbit coupling (SOC) which favors the study of spin related 
phenomena, for instance, spin field-effect transistors,\cite{Chuang2014:NatNanotech} 
and the search for the elusive Majorana fermions.\cite{Das2012:NatPhys} Also, InAs 
WZ shows remarkable piezoelectric and piezoresistive properties\cite{Li2015:AdvMater} 
that, combined with the InAs narrow band gap, can operate in the near-infrared regime. 
On the other hand, InP is a promising candidate for photovoltaic applications\cite{Joyce2013:Nanotech} 
and for enhancing the efficiency of solar cells.\cite{Cui2013:NL} In fact, a silicon-integrated 
nanolaser of InP nanowire has already been demonstrated at room temperature with 
a wide wavelength range due to the WZ/ZB polytypism.\cite{Wang2013:NL} Furthermore, 
it is also possible to combine InAs and InP WZ compounds in axial\cite{Svensson2013:Nanotech} 
and radial\cite{Lindgren2013:Nanotech} heterostructures, which opens the path 
for novel opportunities in band gap engineering.

Theoretically, studies based on WZ III-V compounds including InAs and InP were 
reported using different {\it ab initio} approaches. The main focus of these studies 
was the determination of the lattice parameters, band gaps and SOC energy splittings 
in the valence band.\cite{Zanolli2007:PRB,De2010:PRB,Dacal2011:SSC,Belabbes2012:PRB,Hajlaoui2013:JPdAP,Dacal2014:MatResEx} 
Of these references, De and Pryor\cite{De2010:PRB} provide useful information that 
can be used in effective models, such as the effective masses and the spin splitting 
parameters. The issue of using these parameters is that they are only valid in the 
immediate vicinity of the $\Gamma$ point ($\sim 2 \%$ of the FBZ), limiting the 
range of physical phenomena that can be investigated. In order to achieve a better 
description further away from the $\Gamma$ point, a multiband effective model is 
desirable. Although {\bf k.p} models and parameters are well established for WZ 
III-nitride compounds\cite{Chuang1996:PRB,Rinke2008:PRB}, there are only few reports 
in the literature for non-nitrides, such as InP\cite{FariaJunior2012:JAP,FariaJunior2014:JAP} 
and GaAs.\cite{Cheiwchanchamnangij2011:PRB}
  
In this study, we develop a robust 8$\times$8 {\bf k.p} Hamiltonian to describe 
the {\it ab initio} band structure calculated by WIEN2k\cite{WIEN2k} of InAs and InP in 
WZ phase. We show that our fitted parameters reproduce the {\it ab initio} band structure, 
capturing the important anti-crossings and spin splitting features up to 1.0 nm$^\textrm{-1}$ 
($\sim 10 \%$ of the FBZ in the $k_xk_y$ plane and $\sim 22$\% in the $k_z$ direction). 
At $\Gamma$ point, each band is two-fold degenerate and for the valence band we 
found that the band ordering, from top to bottom, is HH (heavy hole), CH (crystal 
field split-off hole) and LH (light hole) for InAs and HH, LH and CH for InP. 
This ordering is due to an interplay of SOC energy splittings and the crystal field energy 
splitting. The intricate behavior of spin splittings, arising from the bulk inversion 
asymmetry (BIA) of WZ structure, is correctly described by the $k$-dependent SOC 
term, often neglected in the literature. Calculating the spin expectation value 
for the Bloch states, we find a Rashba-like spin texture\cite{Zutic2004:RMP} with 
either clockwise (CW) or counterclockwise (CCW) orientation. All these spin-dependent 
features extracted from our {\bf k.p} Hamiltonian and parameter sets were systematically 
checked to agree with {\it ab initio} calculations. Furthermore, based on our effective 
8$\times$8 Hamiltonian, we calculated the density of states (DOS) and predict 
the carrier density as a function of the Fermi energy. We also provide an analytical 
description of conduction band valid close to $\Gamma$ point and a compact description 
of the valence band (6$\times$6 Hamiltonian). But, we would like to emphasize that 
the best description of InAs and InP WZ is obtained using the total 8$\times$8 
Hamiltonian. {\it In summary, the main goal of our paper is to provide a realistic 
{\bf k.p} description that can be used to study charge and spin transport, optics, 
as well as (superconducting) proximity effects in semiconductor heterostructures, 
e.~g., quantum wells and wires. Such heterostructures cannot be investigated by 
first principles due to their mesoscopic sizes, and {\bf k.p} technique (using the 
prescription $\vec{k} \to -i \vec{\nabla}$) is currently perhaps the best choice 
for obtaining physically relevant quantities for them.}

This paper is organized as follows: in Sec. II we present the {\it ab initio} band 
structure of InAs and InP WZ. The multiband {\bf k.p} model and its considerations 
are discussed in Sec. III. In Sec. IV, we describe our main results: (i) the fitting 
approach; (ii) the comparison between the {\it ab initio} and {\bf k.p} for band 
structure and the spin splittings; (iii) the spin expectation value for all energy 
bands and (iv) the DOS extracted from the 8$\times$8 Hamiltonian. The analytical 
description of CB close to $\Gamma$ point is presented in Sec. V and the compact 
form of valence Hamiltonian, along with its parameters, is shown in Sec. VI. Finally, 
in Sec. VII we present our conclusions and possible direct applications of our effective 
multiband model.

%===============================================================================

\section{{\it ab initio} band structure}
\label{sec:abinitio}

The {\it ab initio} electronic structure for InAs and InP in WZ phase was calculated 
within the density functional theory (DFT) framework,\cite{Hohenberg1964:PR} using 
the full potential linearized augmented plane wave method implemented by the 
WIEN2k code.\cite{WIEN2k} To account for local and semilocal functional deficiencies 
to correctly describe band gaps in semiconductors, we used an efficient and accurate 
alternative for electronic structure calculations based on the modified Becke-Johnson 
(mBJ) exchange potential\cite{Becke2006:JCP} with LDA (local density approximation) 
correlation.\cite{Tran2009:PRL} It has been shown that the semilocal mBJ exchange 
potential provides prediction of band gaps of the same order\cite{Kim2010:PRB,Koller2011:PRB,Koller2012:PRB} 
as hybrid functionals\cite{Kim2009:PRB} and GW method.\cite{Chantis2006:PRL,Luo2009:PRL,Chantis2010:PRB} 
In addition, the semilocal approach to the exchange-correlation functional is barely 
expensive when compared to the LDA\cite{Perdew1981:PRB} or the generalized gradient 
approximation.\cite{Perdew1996:PRL} The SOC is included within the second variational 
step.\cite{Singh2006:book} Regarding the technical details of our calculations, 
we expanded the wave functions in atomic spheres for orbital quantum numbers up to 
10; the plane wave cut-off multiplied with the smallest atomic radii equals to 10 
and the irreducible Brillouin zone was sampled with 600 $k$ points. Further details 
on {\it ab initio} calculations of III-V semiconductors, either with ZB or WZ 
structure, using the mBJ potential can be found in Ref. \onlinecite{Gmitra2016:arXiv}.

\begin{figure}[h!]
\begin{center} 
\includegraphics{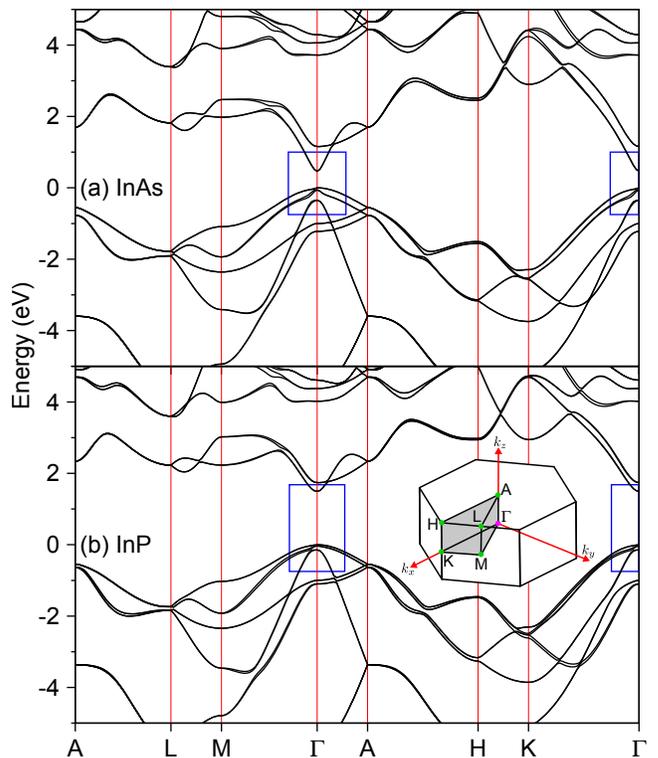}
\caption{(Color online) {\it ab initio} band structure along high symmetry lines for 
(a)~InAs and (b)~InP in WZ phase. The inset shows the FBZ of WZ structure indicating 
the high symmetry points. The rectangles highlight the region of interest around 
the $\Gamma$ point.}
\label{fig:full_bs}
\end{center}
\end{figure}

\begin{figure}[h!]
\begin{center} 
\includegraphics{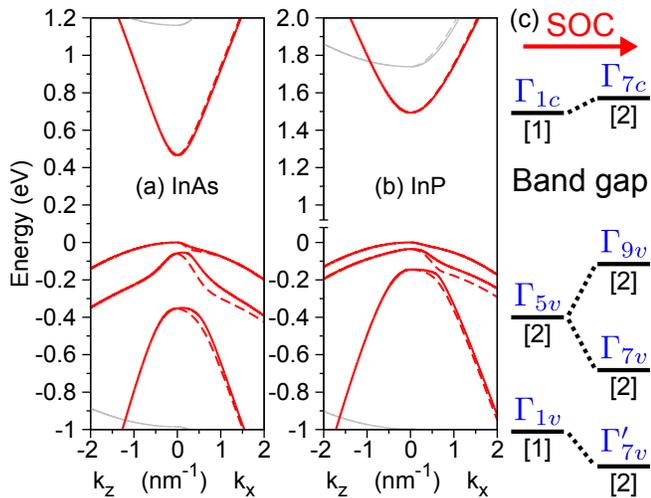}
\caption{(Color online) Band structure for WZ (a) InAs and (b) InP around $\Gamma$ point 
for $k_z$ ($\Gamma$-A) and $k_x$ ($\Gamma$-K) directions. The solid lines indicate 
the outer branch and the dashed lines indicate the inner branch of the spin-split 
bands. The thin (gray) lines indicate the energy bands outside our range of interest. 
(c) Change in the irreducible representations of energy bands at $\Gamma$ point 
under SOC. The subscripts $v$ and $c$ added to the irreducible representations 
indicate valence and conduction bands, respectively, and the prime distinguish between 
the two possibilities of $\Gamma_7$. The numbers in square brackets are the degeneracy 
of the bands. Our notation for the irreducible representations follows Ref.~\onlinecite{Dresselhaus:2008}.}
\label{fig:zoom_bs}
\end{center}
\end{figure}

The particular order of cation (In) and anions (As, P) within the unit cell determines 
spin orientation.\cite{Cardona1988:PRB} We consider the following primitive basis 
vectors for corresponding hexagonal Bravais lattice, $\vec{a}_1=a(\sqrt{3},-1,0)/2$, 
$\vec{a}_2=a(0,1,0)$, and $\vec{a}_3=c(0,0,1)$, where $a$ and $c$ are the 
WZ lattice parameters. Using the three basis vectors $\vec{a}_i$ ($i=1,2,3$) we 
define the following four atomic positions that form the WZ structure: $(2/3,1/3,u)$ 
and $(1/3,2/3,1/2+u)$, with $u=0$ for anion and and $u=3/8$ for cation. We note 
that in general there might be $u=3/8+\epsilon$ with a small dimensionless cell-internal 
structural parameter $\epsilon$ describing a deviation from ideal tetrahedrons as 
one observes for SiC polytypes.\cite{Kackell1994:PRB} In our calculations we 
considered $\epsilon=0$ since it is a rather small valued parameter.\cite{McMahon2005:PRL,Belabbes2012:PRB} 
For the lattice parameters we considered $a=4.2742$~$\mathrm{\AA}$ and 
$c=7.025$~$\mathrm{\AA}$\cite{Kriegner2011:NL} for InAs and $a=4.1148$~$\mathrm{\AA}$ 
and $c=6.7515$~$\mathrm{\AA}$\cite{Panse2011:PRB} for InP.

We show the band structures obtained with WIEN2k in Fig.~\ref{fig:full_bs}(a) for 
InAs and Fig.~\ref{fig:full_bs}(b) for InP. Both compounds show a direct band 
gap at $\Gamma$ point with values of $E_g=0.467\;\textrm{eV}$ for InAs and 
$E_g = 1.494\;\textrm{eV}$ for InP. Due to the hexagonal symmetry of WZ, the $\Gamma$ 
point, as well as the symmetry line connecting $\Gamma$-A (hexagonal axis), belong 
to $C_{6v}$ symmetry group\cite{Casella1959:PR}, that has only two dimensional 
double group representations. From this follows that the states along the hexagonal 
axis are spin degenerate.\cite{Casella1959:PR,Hopfield1961:JAP} Irreducible 
representations of other points in the FBZ compatible with spin are singly degenerate. 
Hence, except for accidental or time-reversal degeneracies at $\Gamma$ and $A$ 
points, spin splittings must occur for all bands.

In Figs.~\ref{fig:zoom_bs}(a), for InAs, and \ref{fig:zoom_bs}(b), for InP, we display 
the rectangular regions of Fig.~\ref{fig:full_bs}, i.~e., a zoom of the band structure 
around $\Gamma$ point. At this energy range, the anti-crossings and spin splittings 
features of the band structures are evident. Because of InAs large SOC, the valence 
band energy levels are further apart than InP bands and additional curvatures are 
present along $k_z$ direction. For InP the top two valence bands along $k_z$ shows 
similar curvatures and no anti-crossing is visible. The effect of SOC in the energy 
bands at $\Gamma$ point is shown schematically in Fig.~\ref{fig:zoom_bs}(c). Without 
SOC, the irreducible representations belong to the simple group, while with SOC, 
they are referred to as double group. This distinction is important for {\bf k.p} 
perturbative approaches.

We present a comparison between our {\it ab initio} calculations and other theoretical 
papers of the literature in table \ref{tab:comparison_theory}. Besides the lattice 
constants $a$ and $c$ we compare the values of the internal parameter $u$, the 
energy gap $E_g$, and the energy difference between the top valence band $\Gamma_{9v}$ 
to the other bands $\Gamma_{7v}$ and $\Gamma^\prime_{7v}$ [following the notation 
of Fig.~\ref{fig:zoom_bs}(c)]. These energy differences are defined as $\Delta E_{97} = E(\Gamma_{9v})-E(\Gamma_{7v})$ 
and $\Delta E^\prime_{97} = E(\Gamma_{9v})-E(\Gamma^\prime_{7v})$. It is very 
common to compare the crystal field splitting energy, $\Delta_1$, and the SOC 
energy, $\Delta_{SO}$, however, these parameters are usually obtained under the 
quasicubic approximation and do not provide a direct comparison with experiments 
such as $\Delta E_{97}$ and $\Delta E_{97^\prime}$. We can see that all the values 
obtained by our calculations are within the range of reported data in previous 
papers. We also compare experimental measurements of the energy gap with our calculated 
values, shown in table \ref{tab:comparison_gaps}. We focused on experimental data 
obtained by photoluminescence measurements at low temperature of large diameter 
nanowires, so that lateral quantum confinement is negligible. For both InAs (despite 
the reduced set of available data) and InP compounds, our calculated values of the 
energy gaps are consistent with the experiments. Furthermore, photoluminescence 
excitation measurements can probe the $\Gamma_{7v}$ and $\Gamma^\prime_{7v}$ valence 
bands and allow us to check our calculated values for $\Delta E_{97}$ and $\Delta E^\prime_{97}$ 
energies. To the best of our knowledge, such experiments are only available for 
InP. Typical values found for $\Delta E_{97}$ and $\Delta E^\prime_{97}$ in InP WZ 
are: 0.044 eV and 0.187 eV in Ref. \onlinecite{Gadret2010:PRB}; 0.043 eV and 0.179 eV 
in Ref. \onlinecite{DeLuca2015:NL}; and 0.044 eV and 0.182 eV in Ref. \onlinecite{Zilli2015:ACSNano}. 
Our calculated values for InP of $\Delta E_{97}=0.0354\;\textrm{eV}$ and 
$\Delta E^\prime_{97}=0.145\;\textrm{eV}$ (from table \ref{tab:comparison_theory}), 
are also in good agreement with these experimental trends. For completeness, we 
provide in appendix A the calculated values of effective masses around $\Gamma$-point.

%Gadret2010:PRB: Ea = 1.488, Eb = 1.532, Ec = 1.675
%DeLuca2015:NL: Ea = 1.494, Eb = 1.537, Ec = 1.673
%Zilli2015:ACSNano: Ea = 1.493, Eb = 1.537, Ec = 1.675

%===============================================================================

\section{\lowercase{k.p} formulation}
\label{sec:kp}

One alternative approach to {\it ab initio} band structure calculations is the {\bf k.p} 
method. In the {\bf k.p} approach, the many-body interactions of electrons with 
nuclei and other electrons are described by an effective potential which has the 
same periodicity as the Bravais lattice of the crystal.\cite{Enderlein1997} Such 
periodic property of the potential allows us to use Bloch's theorem for the total 
wave function. The single-particle Hamiltonian for the periodic part of the Bloch 
function, $u_{n,\vec{k}}(\vec{r})$, can be written as:
\begin{eqnarray}
\mathbf{H} & = & \underbrace{\frac{p^{2}}{2m_{0}}+V(\vec{r})}_{\mathbf{H_{0}}}\,+\, \underbrace{\frac{\hbar}{4m_{0}^{2}c^{2}}\left[\vec{\nabla}V(\vec{r})\times\vec{p}\right]\cdot\vec{\sigma}}_{\mathbf{H_{SO}}} \nonumber \\
& + & \underbrace{\frac{\hbar^{2}k^{2}}{2m_{0}}}_{\mathbf{H_{k2}}} + \underbrace{\frac{\hbar}{m_{0}}\vec{k}\cdot\vec{p}}_{\mathbf{H_{kp}}} \, + \, \underbrace{\frac{\hbar^{2}}{4m_{0}^{2}c^{2}}\left[\vec{\nabla}V(\vec{r})\times\vec{k}\right]\cdot\vec{\sigma}}_{\mathbf{H_{kSO}}} \, ,
\label{eq:Hunk}
\end{eqnarray}
in which the different terms in the Hamiltonian are identified for convenience.

We can solve the above equation perturbatively expanding the functions $u_{n,\vec{k}}(\vec{r})$ 
around a specific reciprocal space point that we know the solutions for the Hamiltonian. 
Since WZ InAs and InP have a direct band gap at $\Gamma$ point, this is the 
chosen expansion point. The perturbative technique we use in this paper is L\"owdin's 
formalism.\cite{Lowdin1951:JCP} In this approach, the functions at $\Gamma$ point, 
i.~e., the basis set to expand $u_{n,\vec{k}}(\vec{r})$, are divided into classes 
A and B. The energy bands we are interested in describing comprise the class A while 
the other energy bands belong to class B. The contribution of states in class B appear 
in second or higher orders of perturbation. The matrix elements we consider can 
arise from first or second order perturbation, reading as
\begin{equation}
H_{f,\alpha\alpha'}^{(1)} = \left\langle \alpha \left| \mathbf{H_{f}} \right| \alpha'\right\rangle \, ,
\end{equation}
and
\begin{equation}
H_{fg,\alpha\alpha'}^{(2)} = \sum_{\beta}^{B}\frac{\left\langle \alpha\left| \mathbf{H_{f}} \right|\beta\right\rangle \left\langle \beta\left| \mathbf{H_{g}} \right|\alpha'\right\rangle }{E_{\alpha\alpha'}-E_{\beta}} \, ,
\end{equation}
where $\mathbf{H_{f}}$ and $\mathbf{H_{g}}$ can be any of the terms of equation (\ref{eq:Hunk}), 
except $\mathbf{H_{0}}$.

%-------------------------------------------------------------------------------
% tables from sec AB INITIO BAND STRUCTURE
\onecolumngrid

\begin{table}[H]
\caption{Comparison of theoretical data for InAs and InP in WZ phase. The lattice 
constants $a$ and $c$ are given in $\textrm{\AA}$ and $u$ is dimensionless. The 
band gap, $E_g$, and the valence band energy differences, $\Delta E_{97}$ and 
$\Delta E^\prime_{97}$, are given in eV.}
\begin{center}
{\renewcommand{\arraystretch}{1.2}
\begin{tabular*}{1\columnwidth}{@{\extracolsep{\fill}}llcccccc}
\hline
\hline 
 &  & $a$ & $c$ & $u$ & $E_{g}$ & $\Delta E_{97}$ & $\Delta E_{97}^{\prime}$\tabularnewline
\hline 
InAs & This study                               & 4.2742 & 7.0250 & 0.37500 & 0.4670 & 0.0592 & 0.3527\tabularnewline
     & Ref. \onlinecite{De2010:PRB}             & 4.1505 & 6.7777 & 0.37500 & 0.4810 & 0.1050 & 0.4690\tabularnewline
     & Ref. \onlinecite{Belabbes2012:PRB}$^{*}$ & 4.2570 & 6.9894 & 0.37447 & 0.4810 & 0.0573 & 0.3937\tabularnewline
     & Ref. \onlinecite{Dacal2014:MatResEx}     & 4.2564 & 7.0046 & 0.37400 & 0.4610 & 0.0700 & 0.3640\tabularnewline
     & Ref. \onlinecite{Gmitra2016:arXiv}       & 4.2742 & 7.0250 & 0.37422 & 0.4610 & 0.0660 & 0.3600\tabularnewline
\hline 
InP & This study & 4.1148 & 6.7515             & 0.37500 & 1.4940 & 0.0354 & 0.1450 \tabularnewline
    & Ref. \onlinecite{De2010:PRB}             & 4.2839 & 6.9955 & 0.37500 & 1.4740 & 0.0630 & 0.3480 \tabularnewline
    & Ref. \onlinecite{Dacal2011:SSC}          & 4.1500 & 6.9120 & 0.37100 & 1.4936 & 0.0450 & 0.2430 \tabularnewline
    & Ref. \onlinecite{Belabbes2012:PRB}$^{*}$ & 4.1148 & 6.7515 & 0.37458 & 1.5760 & 0.0321 & 0.1339 \tabularnewline
\hline
\hline
\multicolumn{8}{l}{$^{*}$$a$, $c$ and $u$ from Ref. \onlinecite{Panse2011:PRB}}\tabularnewline
\end{tabular*}}
\end{center}
\label{tab:comparison_theory}
\end{table}

\twocolumngrid

\begin{table}[h!]
\caption{Comparison between theoretical and experimental values of the energy gap. 
We indicate the temperature of the photoluminescence measurements in parenthesis.}
\begin{center}
{\renewcommand{\arraystretch}{1.2}
\begin{tabular*}{1\columnwidth}{@{\extracolsep{\fill}}lcc}
\hline
\hline
      & $E_{g}$ (eV) & $E_{g}$ (eV) \tabularnewline
      & This study   & Experiment   \tabularnewline
\hline
InAs  & 0.467        & 0.520 (7 K)$^{a}$, 0.500 (20 K)$^{b}$ \tabularnewline
      &              & 0.458 (5 K)$^{c}$   \tabularnewline
\hline
InP  & 1.494         & 1.492 (10 K)$^{d}$, 1.494 (10 K)$^{e}$ \tabularnewline
     &               & 1.490 (20 K)$^{f}$, 1.491 (4 K)$^{g}$  \tabularnewline
     &               & 1.493 (4 K)$^{h}$,  1.488 (6 K)$^{i}$   \tabularnewline
\hline
\hline
\multicolumn{3}{l}{ $^{a}$Ref. \onlinecite{Bao2009:AM}, $^b$Ref. \onlinecite{Koblmuller2010:Nanotech}, 
$^{c}$Ref. \onlinecite{Moller2012:Nanotech}, $^{d}$Ref. \onlinecite{Zilli2015:ACSNano}, 
$^{e}$Ref. \onlinecite{DeLuca2015:NL}, }\tabularnewline
\multicolumn{3}{l}{$^{f}$Ref. \onlinecite{Mishra2007:APL}, $^{g}$Ref. \onlinecite{Vu2013:Nanotech}, 
$^{h}$Ref. \onlinecite{Tuin2011:NanoRes}, $^{i}$Ref. \onlinecite{Gadret2010:PRB}}\tabularnewline
\end{tabular*}}
\end{center}
\label{tab:comparison_gaps}
\end{table}
%-------------------------------------------------------------------------------

Since the unperturbed term, $\mathbf{H_{0}}$, in Eq.~(\ref{eq:Hunk}) does not contain 
SOC effects explicitly, we consider the simple group description of the energy bands, 
the most usual approach in the literature.\cite{note:doublegroup} Under such approximation, 
the states in class A belong to the irreducible representations shown in the left 
side of Fig.~\ref{fig:zoom_bs}(c), a 4 dimensional Hilbert space, combined with 
the spin 1/2 angular momentum, a 2 dimensional Hilbert space. Therefore, the 8 
dimensional basis set for the {\bf k.p} Hamiltonian in Dirac notation\cite{note:dirac} 
is given by:
\begin{equation}
\begin{aligned}
\left|c_{1}\right\rangle & = -\frac{\left|(\Gamma_{5v}^x+i\Gamma_{5v}^y)\uparrow\right\rangle}{\sqrt{2}} & \\
\left|c_{2}\right\rangle & = \frac{\left|(\Gamma_{5v}^x-i\Gamma_{5v}^y)\uparrow\right\rangle}{\sqrt{2}}  & \\
\left|c_{3}\right\rangle & = \left|\Gamma_{1v}\uparrow\right\rangle &  \\
\left|c_{4}\right\rangle & = \frac{\left|(\Gamma_{5v}^x-i\Gamma_{5v}^y)\downarrow\right\rangle}{\sqrt{2}}  & \\
\end{aligned}
\begin{aligned}
\left|c_{5}\right\rangle & = -\frac{\left|(\Gamma_{5v}^x+i\Gamma_{5v}^y)\downarrow\right\rangle}{\sqrt{2}} & \\
\left|c_{6}\right\rangle & = \left|\Gamma_{1v}\downarrow\right\rangle & \\
\left|c_{7}\right\rangle & = i\left|\Gamma_{1c}\uparrow\right\rangle & \\
\left|c_{8}\right\rangle & = i\left|\Gamma_{1c}\downarrow\right\rangle \, , & \\
\end{aligned}
\label{eq:cbasis}
\end{equation}
with 1-6 representing the valence band states and 7-8 the conduction band states. 
Since $\Gamma_{5v}$ is two dimensional, we identified its basis states by $\left| \Gamma_{5v}^x \right\rangle \sim x$ 
and $\left| \Gamma_{5v}^y \right\rangle \sim y$. The single arrows ($\uparrow,\downarrow$) 
represent the projection of spin up and spin down, eigenvalues of $\sigma_z$ Pauli 
matrix. The states in class B have simple group symmetries $\Gamma_1$, $\Gamma_3$, 
$\Gamma_5$ and $\Gamma_6$, which is the only necessary information to calculate 
second order contributions.

\begin{figure}[h!]
\begin{center} 
\includegraphics{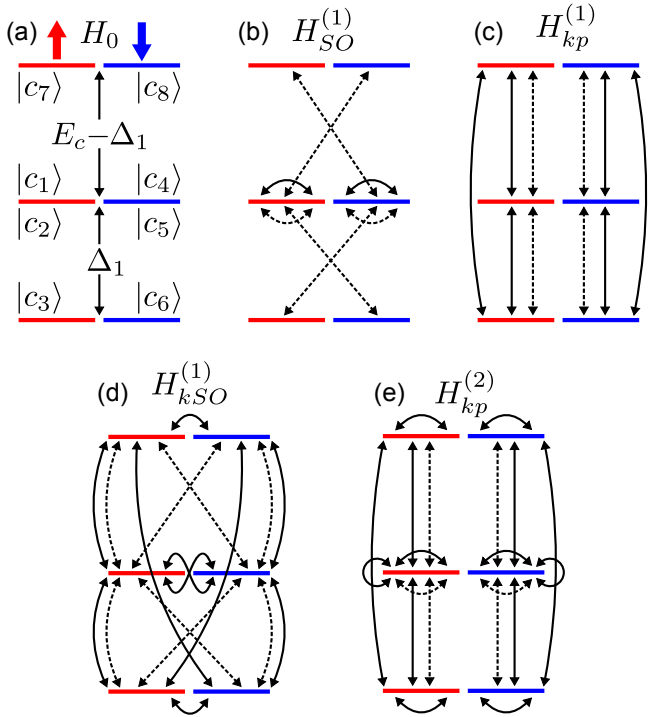}
\caption{(Color online) Possible interactions of the Hamiltonian terms (a) $H_0$, 
(b) $H^{(1)}_{SO}$, (c) $H^{(1)}_{kp}$, (d) $H^{(1)}_{kSO}$ and (e) $H^{(2)}_{kp}$. 
The arrows on top of panel (a) indicate spin up and spin down projections of the 
basis states. Since $\left|c_{1(4)}\right\rangle$ and $\left|c_{2(5)}\right\rangle$ 
are degenerate, we indicate the interactions arising from $\left|c_{1(4)}\right\rangle$ 
with solid lines and the interactions arising from $\left|c_{2(5)}\right\rangle$ 
with dashed lines. For the other states without degeneracy we used solid lines. 
In panel (a) we show the energy splittings without SOC, formally defined in appendix B.}
\label{fig:kp_terms}
\end{center}
\end{figure}

To describe the interaction among the energy bands, we consider all terms of 
equation (\ref{eq:Hunk}) in first order perturbation and only the term $\mathbf{H_{kp}}$ 
in second order. Therefore, the total matrix Hamiltonian in the basis set 
(\ref{eq:cbasis}) comprises the following terms
\begin{equation}
H = H_0 + H^{(1)}_{SO} + H^{(1)}_{kp} + H^{(1)}_{kSO} + H^{(2)}_{kp} \, ,
\label{eq:Hmatrix}
\end{equation}
with the explicit form of each matrix and the definition of the parameters given 
in the appendix B.

In Fig.~\ref{fig:kp_terms} we show schematically the interactions for each term 
in the total Hamiltonian (\ref{eq:Hmatrix}). The panel~\ref{fig:kp_terms}(a) represents 
the unperturbed Hamiltonian without SOC, where states $\left|c_{1(4)}\right\rangle$ 
and $\left|c_{2(5)}\right\rangle$ are degenerate for spin up (down). The only terms 
that couple different spin projections arise from $H^{(1)}_{SO}$ or $H^{(1)}_{kSO}$, 
panels~\ref{fig:kp_terms}(b) and~\ref{fig:kp_terms}(d), respectively. Usually $H^{(1)}_{kSO}$ 
is neglected in WZ Hamiltonians\cite{Chuang1996:PRB,Beresford2004:JAP,Fu2008:JAP,Rinke2008:PRB,
Marnetto2010:JAP,Cheiwchanchamnangij2011:PRB,Miao2012:PRL}. However, the explicit 
interactions for non-zero $k$-values are crucial to correctly describe the spin 
splitting properties. We included $H^{(1)}_{kSO}$ following the approach of 
Dresselhaus for ZB~\cite{Dresselhaus1955:PR}. Moreover, the coupling of $H^{(1)}_{SO}$ 
to other terms provides additional contributions to the spin splitting of energy bands. 
Besides spin splitting properties, we want a good description of the band structure 
curvatures. Such effects can be modeled by linear and quadratic terms of the $H^{(1)}_{kp}$ 
and $H^{(2)}_{kp}$, panels~\ref{fig:kp_terms}(c) and~\ref{fig:kp_terms}(e), respectively. 
The only term that allows a $k$-dependent self interaction of states is $H^{(2)}_{kp}$ 
which gives the effective mass contribution to our model. 

Although the {\bf k.p} method provides the functional form of the Hamiltonian, 
the parameters that describe different materials cannot be found by group theory 
arguments only. In order to calculate the matrix elements we would need the functions 
at the expansion point and also the periodic potential $V(\vec{r})$. Alternatively, 
we can directly fit the {\bf k.p} Hamiltonian to the {\it ab initio} band structure to 
extract the parameters.\cite{Beresford2004:JAP,Rinke2008:PRB,Marnetto2010:JAP,Cheiwchanchamnangij2011:PRB,Punya2012:PRB}

%===============================================================================

\section{Numerical fitting of the 8$\times$8 \lowercase{k.p} Hamiltonian}
\label{sec:fit8}

We start our fitting approach by calculating the $k$-independent parameters of the 
Hamiltonian, i.~e., the energy splittings. The values for crystal field splitting, 
$\Delta_1$, and the conduction band energy, $E_c$, can be obtained from the {\it ab initio} 
calculation without SOC, which is in fact the assumption of the {\bf k.p} perturbative 
theory [$H_0$ term, see Fig~\ref{fig:kp_terms}(a)]. This approach is very useful 
because it simplifies the calculation of the SOC energy splittings inside valence 
band, $\Delta_2$ (coupling same spins) and $\Delta_3$ (coupling different spins), 
and the SOC between conduction and valence bands, $\Delta_4$. Please refer to appendix 
B for the formal definition of these splitting energies. By setting the values 
of $\Delta_1$ and $E_c$, it possible to have $\Delta_2 \neq \Delta_3$ and neglect 
the cubic approximation.\cite{Chuang1996:PRB} If the values of $\Delta_1$ and 
$E_c$ were not found without SOC, we would have to determine 5 variables having 
only 3 linear independent combinations of the energy bands with SOC. This approach 
would provide a range of possible values and further analysis would be necessary. 
Starting with $\Delta_1$ and $E_c$ values without SOC, we obtained four different 
solution sets for the SOC splitting energies since $\Delta_3$ and $\Delta_4$ are 
off-diagonal terms in the Hamiltonian and can assume positive or negative values 
with same magnitude. At $\Gamma$ point any of these solution sets give the same 
eigenvalues, therefore we set $\Delta_3$ to be positive\cite{De2010:PRB,Dacal2011:SSC,Belabbes2012:PRB,Dacal2014:MatResEx} 
and investigated the effect of positive and negative values of $\Delta_4$.

\begin{figure}[h!]
\begin{center} 
\includegraphics{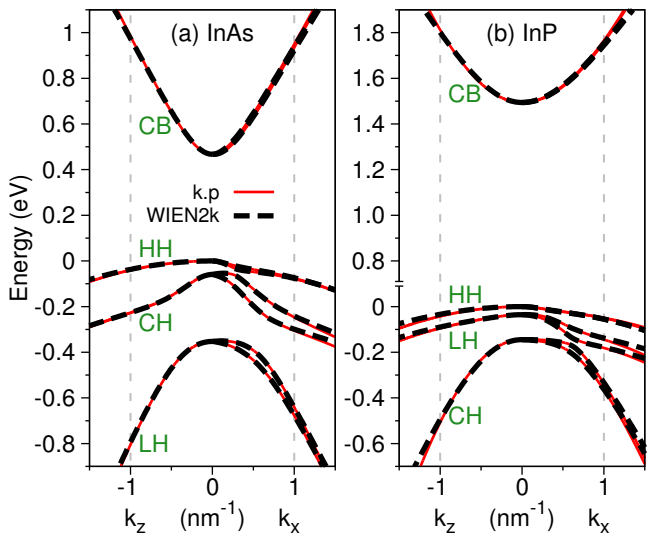}
\caption{(Color online) Comparison of band structures calculated from the fitted 
{\bf k.p} model (solid lines) and the {\it ab initio} WIEN2k (dashed lines)
for (a) InAs and (b) InP. The vertical dashed lines at 1.0 nm$^\textrm{-1}$ indicate 
the borders of the fitting range.}
\label{fig:fitted_bs}
\end{center}
\end{figure}

Before starting the fitting of the $k$-dependent parameters, it is important to 
define the fitting region we are interested in, which is connected to the limits of 
our {\bf k.p} model. Basically, in order to describe as precisely as possible the 
8 bands we are interested in, we should stay in a region away from the influence of 
remote bands, roughly $k \sim 1.5$ nm$^\textrm{-1}$, see Fig.~\ref{fig:zoom_bs}(a)-(b). 
We also want to have a nice description of the anti-crossings in the bands structure 
around $k \sim 0.5$ nm$^\textrm{-1}$. Furthermore, in the {\bf k.p} Hamiltonian $k_x$ 
and $k_y$ directions are equivalent, but this is not the case for the {\it ab initio} 
band structure. Around $k \sim 1.0$ nm$^\textrm{-1}$, the {\it ab initio} band structures 
along $\Gamma$-K and $\Gamma$-M directions are different, especially the spin 
splitting, which is another feature to be described. Therefore, it is reasonable 
to set the goal of our fitting at $k = 1.0$ nm$^\textrm{-1}$ to find the best 
parameters set that describes the {\it ab initio} band structure around $\Gamma$ point 
for all the 8 bands.

To increase the accuracy of our parameter sets, we fitted, simultaneously, the 
energy bands in multiple directions of the FBZ ($\Gamma$-K, $\Gamma$-M, $\Gamma$-A, 
$\Gamma$-H and $\Gamma$-L). The fitting algorithm was developed using the LMFIT\cite{lmfit}
package of python assuming several minimization methods available. We noticed that 
the minimization methods behave differently and usually provide different parameter 
sets. After an initial fit, we chose the best parameter set and used it as input 
for a new fit using all minimization methods again. To find the best fit, the band 
structures and spin splittings are compared by their residue\cite{note:residue} 
up to $k = 1.0$ nm$^\textrm{-1}$ for all directions. The best parameter sets 
for InAs and InP found by our fitting approach are presented in table~\ref{tab:par_full}.

\begin{figure}[h!]
\begin{center} 
\includegraphics{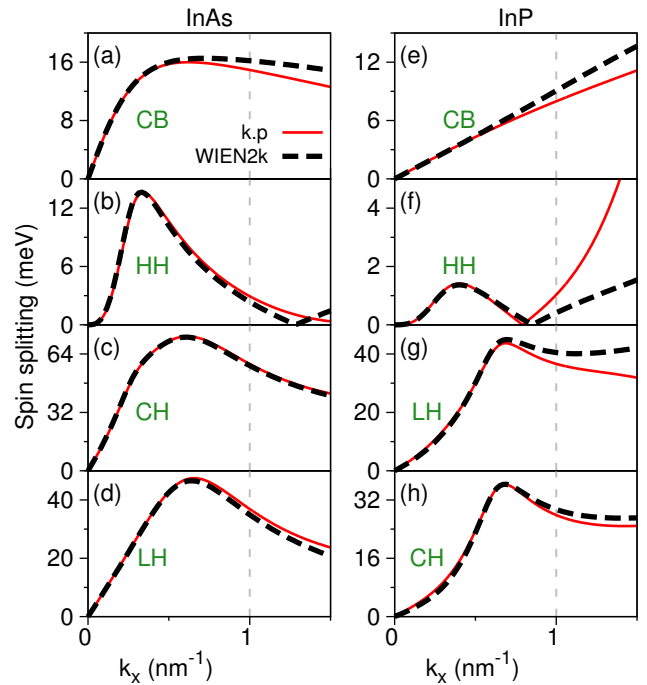}
\caption{(Color online) Comparison of the spin splittings along $k_x$ for the energy 
bands (a) CB, (b) HH, (c) CH and (d) LH of InAs and (e) CB, (f) HH, (g) LH and 
(h) CH of InP. The line schemes follow Fig.~\ref{fig:fitted_bs}.}
\label{fig:fitted_sp_GK}
\end{center}
\end{figure}

In Fig.~\ref{fig:fitted_bs}, we present the comparison between the fitted and WIEN2k 
{\it ab initio} band structures along $k_z$ and $k_x$ for InAs and InP. All the important 
features around $\Gamma$ point, i.~e., anti-crossings and spin splittings, are 
captured by our model. We notice a good agreement up to $k = 1.0$ nm$^\textrm{-1}$ 
with small deviations above it, indicating that we are reaching the region where 
the influence of remote energy bands becomes important. We labeled the valence bands 
according to the composition of states at $\Gamma$ point. Following Chuang and 
Chang's notation,\cite{Chuang1996:APL} HH is purely composed of $\left|c_{1(4)}\right\rangle$ 
states, LH has more contribution from $\left|c_{2(5)}\right\rangle$ than $\left|c_{3(6)}\right\rangle$ 
states, and CH has more contribution from $\left|c_{3(6)}\right\rangle$ than $\left|c_{2(5)}\right\rangle$ 
states. Since this analysis is usually performed without $\Delta_4$ parameter, we also 
calculated $\Delta_2$ and $\Delta_3$ considering $\Delta_4 = 0$ and we found that 
the same labeling holds (this values are shown in Sec.~\ref{sec:compactVB}). Furthermore, 
we also compared the {\bf k.p} composition with the projection to atomic orbitals 
of the {\it ab initio} wave functions and the same trends can be noticed. The labeling 
order of CH-LH in InAs is due to the values of SOC splitting energies, which are 
slightly larger than the crystal field splitting. For InP, the crystal field splitting 
is dominant leading to LH-CH ordering. Although this labeling of the valence band 
can be confusing, it is very useful to extract optical trends from the band-egde 
transitions. For instance, if we take into account optical transitions arising from 
the top two valence bands, we can expect InP light polarization to be more in-plane 
due to LH contribution than InAs due to CH contribution. Finally, for the conduction 
band of both InAs and InP we simply label it CB, short notation for conduction 
band; CB is mainly composed of $\left|c_{7(8)}\right\rangle$ states.

\begin{figure}[h!]
\begin{center} 
\includegraphics{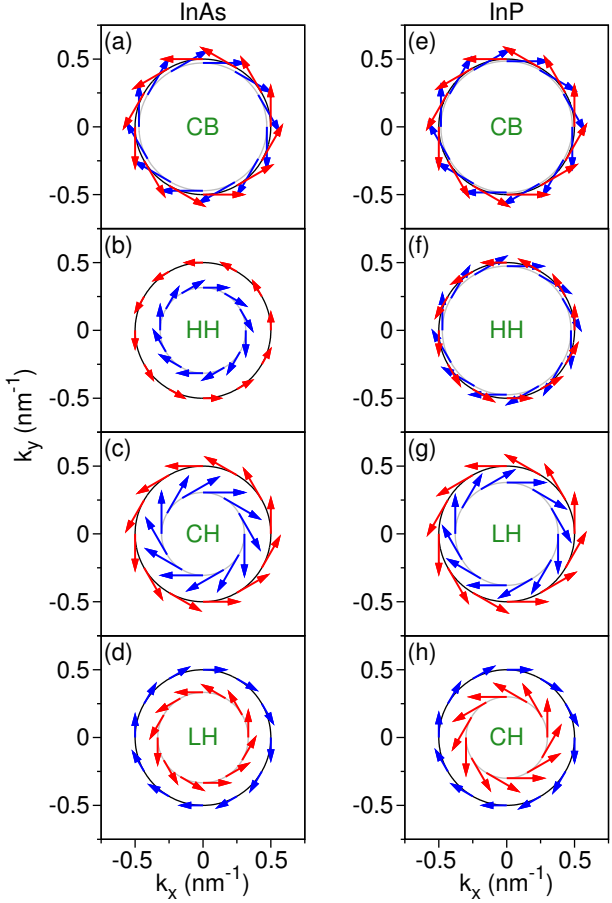}
\caption{(Color online) Spin texture in the $k_xk_y$ plane ($k_z=0$) for the energy 
bands (a) CB, (b) HH, (c) CH and (d) LH of InAs and (e) CB, (f) HH, (g) LH and 
(h) CH of InP. The blue arrows indicate clockwise orientation while the red arrows 
indicate counter clockwise orientation. The amplitude of all arrows are multiplied 
by 0.3 to fit the figure. The constant-energy contours are also drawn in the figure 
in black for the outer branches and in gray for the inner branches.}
\label{fig:spin_orientation}
\end{center}
\end{figure}

\begin{table}
\caption{Parameter sets of the 8$\times$8 Hamiltonian for InAs and InP WZ. The 
energy splittings are given in eV, linear parameters in eV$\cdot\textrm{\AA}$ and 
second order parameters in units of $\hbar^2/2m_0$.}
\begin{center}
{\renewcommand{\arraystretch}{1.2}
\begin{tabular*}{1\columnwidth}{@{\extracolsep{\fill}}ccc}
\hline
\hline
Parameter & InAs & InP \tabularnewline
\hline
Energy splittings &         &        \tabularnewline
$\Delta_{1}$      & 0.1003  & 0.0945 \tabularnewline
$\Delta_{2}$      & 0.1023  & 0.0279 \tabularnewline
$\Delta_{3}$      & 0.1041  & 0.0314 \tabularnewline
$\Delta_{4}$      & 0.0388  & 0.0411 \tabularnewline
$E_{c}$           & 0.6649  & 1.6142 \tabularnewline
\hline
Linear parameters &         &         \tabularnewline
$A_{7}$           & -0.4904 & -0.1539 \tabularnewline
$P_{1}$           &  8.3860 &  7.6349 \tabularnewline
$P_{2}$           &  6.8987 &  5.5651 \tabularnewline
$\alpha_{1}$      & -0.0189 &  0.2466 \tabularnewline
$\alpha_{2}$      & -0.2892 & -0.2223 \tabularnewline
$\alpha_{3}$      & -0.5117 & -0.2394 \tabularnewline
$\beta_{1}$       & -0.0695 & -0.0481 \tabularnewline
$\beta_{2}$       & -0.2171 & -0.1386 \tabularnewline
$\gamma_{1}$      &  0.5306 &  0.2485 \tabularnewline
\hline
Second order parameters &   &         \tabularnewline
$A_{1}$           &  1.5726 & -1.0419 \tabularnewline
$A_{2}$           & -1.6521 & -0.9645 \tabularnewline
$A_{3}$           & -2.6301 & -0.0694 \tabularnewline
$A_{4}$           &  0.5126 & -1.2760 \tabularnewline
$A_{5}$           &  0.1172 & -1.1024 \tabularnewline
$A_{6}$           &  1.3103 & -0.5677 \tabularnewline
$e_{1}$           & -3.2005 & -0.5732 \tabularnewline
$e_{2}$           &  0.6363 &  2.4084 \tabularnewline
$B_{1}$           & -2.3925 & -7.7892 \tabularnewline
$B_{2}$           &  2.3155 &  4.3981 \tabularnewline
$B_{3}$           & -1.7231 &  9.1120 \tabularnewline
\hline
\hline
\end{tabular*}}
\end{center}
\label{tab:par_full}
\end{table}

Let us take a closer look at the spin splitting properties obtained from the {\bf k.p} 
model and the {\it ab initio}. We show the comparison between the two methods in 
Fig.~\ref{fig:fitted_sp_GK} for InAs and InP along $k_x$ direction. Similar to the 
band structure, we have a good agreement up to $k = 1.0$ nm$^\textrm{-1}$ with 
deviations above this region. The intricate behaviors, i.e., the appearance of 
maxima and crossings between HH spin split bands are also described by our model. 
All these spin splitting characteristics have only one physical origin, the BIA 
of WZ structure. The strength of SOC is greater in InAs than InP, visible at the 
peak values and positions. From the largest to the smallest values of the spin 
splitting, we have CH (LH), LH (CH), CB and HH for InAs (InP). Furthermore, a linear 
behavior is maintained for InP CB throughout the fitting region. For InAs, this 
linear behavior is attained only in a small region close to $\Gamma$ point. In appendix 
C, we present the band structure and spin splittings for the other FBZ directions 
used in the fitting.

Another feature we investigated is the spin orientation, i.~e., the spin expectation 
value, $\left\langle \vec{\sigma} \right\rangle$, for the different energy bands, 
presented in Fig.~\ref{fig:spin_orientation} for the $k_xk_y$ plane ($k_z=0$). We 
chose the constant-energy contours to be $E_n(k_x=0.5\;\textrm{nm}^\textrm{-1},k_y=0,k_z=0)$ 
of the outer branch, i.~e., $E_{\textrm{CB}} \sim 630.0$ meV, $E_{\textrm{HH}} \sim -37.2$ 
meV, $E_{\textrm{CH}} \sim -123.0$ meV, $E_{\textrm{LH}} \sim -391.8$ meV for InAs 
and $E_{\textrm{CB}} \sim 1563.5$ meV, $E_{\textrm{HH}} \sim -21.9$ meV, $E_{\textrm{LH}} \sim -75.0$ 
meV, $E_{\textrm{CH}} \sim -156.7$ meV for InP. We found that all the investigated 
energy bands show a Rashba-like spin texture. For InAs, the bands CB, HH and CH 
have the same spin texture, i.~e., CW (CCW) orientation for the inner (outer) branch, 
while LH has the CWW (CW) orientation for the inner (outer) branch. In other words, 
the top two valence bands have the same spin texture while the third valence band 
have the opposite. For InP, the same spin texture holds, even though the labeling 
of CH and LH is reversed. The spin textures calculated with the {\bf k.p} model 
were also checked with the {\it ab initio} calculations.

Performing the fitting approach with the negative sign of $\Delta_4$ we obtained 
the same behavior of the band structure and the spin splittings, but with a reversed 
orientation in the spin texture, i.~e., CW orientation becomes CCW and vice-versa 
for all bands. Specifically, we found that starting with negative value of $\Delta_4$, 
the signs of parameters $A_7$, $\alpha_1$, $\alpha_2$, $\alpha_3$, $\gamma_1$, $B_1$, 
$B_2$ and $B_3$ are changed, but not their amplitude. This change in the spin texture 
is a feature expected from {\it ab initio} regarding the cation and anion positions 
within the crystal unit cell\cite{Cardona1988:PRB} and it is reflected in our {\bf k.p} 
model and parameters. Therefore, in order to provide reliable parameter sets for 
{\bf k.p} Hamiltonians, not only the band structure and the spin splittings should be 
checked but also the spin orientation. We would like to emphasize that all these 
features were systematically checked in this study.

%-------------------------------------------------------------------------------

\subsection{Density of states and carrier density}
\label{sec:DOS_n}

\begin{figure}[h!]
\begin{center} 
\includegraphics{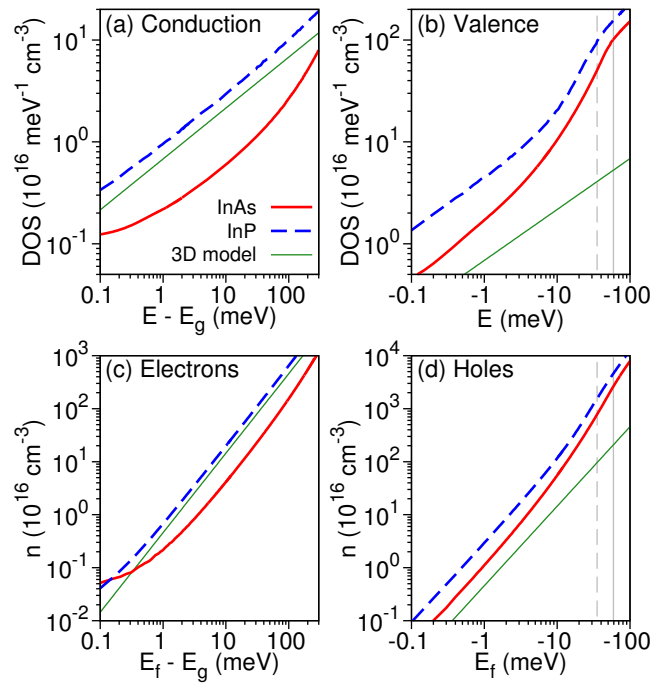}
\caption{(Color online) Calculated DOS for (a) conduction band and (b) valence band 
of InAs, InP and the 3D parabolic band model using an effective mass of $m^*=0.1$. 
Carrier density, $n$, as function of Fermi energy, $E_f$, for (c) electrons and (d) 
holes obtained by the integration of the DOS in Figs (a) and (b), respectively. The 
dashed vertical lines in Figs. (b) and (d) indicate the LH energy at $\Gamma$-point 
for InP while the solid vertical lines indicate the CH energy energy at $\Gamma$-point 
for InAs.}
\label{fig:DOS_logn}
\end{center}
\end{figure}

Relying on the effective 8$\times$8 {\bf k.p} Hamiltonian, it is straightforward 
to calculate a smooth DOS using a fine 3-dimensional (3D) mesh of $k$ points ($300 
\times 300 \times 300$) without much computational effort. In Fig.~\ref{fig:DOS_logn}(a) 
we show the DOS for the conduction band of InAs and InP. For comparison, we also show 
the DOS for the 3D parabolic band model [$\textrm{DOS}(E) \propto \sqrt{E}$], which 
is just a straight line in the log-log scale. Due to the complex behavior of the InAs 
and InP conduction bands, we clearly see deviations from the linear behavior, 
especially for InAs. For the DOS of the valence band, presented in Fig.~\ref{fig:DOS_logn}(b) 
the deviations from the parabolic model are much more visible, showing explicitly 
the need of a multiband approach. When the valence band energy approaches the CH 
(LH) region of InAs (InP), the DOS changes its curvature. Moreover, the valence 
band DOS is approximately one order of magnitude larger than the DOS of the conduction 
band, a behavior attributed to the small curvatures of the valence bands, i.~e., large 
effective masses for holes (in a single band picture). Integrating the DOS we obtain 
the carrier density as a function of the Fermi energy, presented in Figs.~\ref{fig:DOS_logn}(c) 
and (d) for electrons and holes, respectively. Typically, InP supports larger values 
of the carrier density than InAs. For instance, for 100 meV above the energy gap 
$\sim1.6\times10^{18} \; \textrm{cm}^{\textrm{-3}}$ for InAs and $\sim6.5\times10^{18} \; \textrm{cm}^{\textrm{-3}}$. 
In the supplemental material we provide a curve fitting of the carrier density 
curves that can be directly applied to predict the carrier concentration or the 
Fermi energy without the explicit DOS calculation using the 8$\times$8 {\bf k.p} 
Hamiltonian.

%===============================================================================

\section{Analytical description for conduction band}
\label{sec:analyticalCB}

Since the conduction band has a predominant contribution of $\left|c_{7(8)}\right\rangle$ 
states, it is useful to provide an analytical description that holds for small 
regions close to the $\Gamma$ point that can be easily used in spin dynamics 
studies. We apply the L\"owdin's approach again, but now dividing the basis states 
A (Eq. \ref{eq:cbasis}) of the full matrix into new two classes A$^\prime$ ($\left|c_{7,8}\right\rangle$) 
and B$^\prime$ ($\left|c_{1,\cdots,6}\right\rangle$. Using only the terms we already 
calculated in the full Hamiltonian as contribution to the effective Hamiltonian, 
this L\"owdin's approach is usually refereed to as folding down.\cite{Lowdin1951:JCP,Fu2008:JAP} 
The effective Hamiltonian for the first order folding down, keeping terms up to 
$k^3$ can be written as
\begin{equation}
H_{\textrm{CB}} = M(\vec{k})\mathcal{I}_{2}+\vec{\Omega}(\vec{k})\cdot\vec{\sigma} \, ,
\end{equation}
in which $\mathcal{I}_2$ is a $2\times2$ identity matrix and $M$ is the effective mass term 
given by 
\begin{equation}
M = E_{g}+m_{z}k_{z}^{2}+m_{xy}\left(k_{x}^{2}+k_{y}^{2}\right) \, ,
\end{equation}
with the coefficients $m_z$ and $m_{xy}$ given by
\begin{eqnarray}
m_z & = & e_{1} + \frac{P_{1}^{2}}{E_{c}}+\frac{2\beta_{1}^{2}}{E_{c}-\Delta_{1}+\Delta_{2}}+\frac{2\Delta_{4}^{2}\left(A_{1}+A_{3}\right)}{\left(E_{c}-\Delta_{1}+\Delta_{2}\right)^{2}} \, ,\nonumber \\
m_{xy} & = & e_{2} + \frac{1}{2}\frac{\left(P_{2}+\beta_{1}\right)^{2}}{E_{c}-\Delta_{1}+\Delta_{2}}+\frac{1}{2}\frac{\left(P_{2}-\beta_{1}\right)^{2}}{E_{c}-\Delta_{1}-\Delta_{2}} \nonumber \\
       & & +\frac{\beta_{2}^{2}}{E_{c}}+\frac{2\Delta_{4}^{2}\left(A_{2}+A_{4}\right)}{\left(E_{c}-\Delta_{1}+\Delta_{2}\right)^{2}} \, .
\end{eqnarray}

The SOC field $\vec{\Omega}(\vec{k})$ is written as
\begin{equation}
\vec{\Omega}(\vec{k}) = \left[ \alpha +\gamma_{z}k_{z}^{2}+\gamma_{xy}\left(k_{x}^{2}+k_{y}^{2}\right) \right] \left[\begin{array}{c}
k_{y}\\
-k_{x}\\
0
\end{array}\right] \, ,
\end{equation}
with linear and cubic coefficients given by
\begin{eqnarray}
\alpha & = & -\gamma_{1} + \frac{2\Delta_{4}\left(P_{2}+\beta_{1}\right)}{E_{c}-\Delta_{1}+\Delta_{2}} \, , \nonumber \\
\gamma_z & = & 2\sqrt{2}\beta_{1}B_{3}-\frac{2\beta_{2}B_{1}}{E_{c}}+\frac{2\Delta_{4}\left(P_{2}+\beta_{1}\right)\left(A_{1}+A_{3}\right)}{\left(E_{c}-\Delta_{1}+\Delta_{2}\right)^{2}} \, , \nonumber \\
\gamma_{xy} & = &  -\frac{2\beta_{2}B_{2}}{E_{c}}+\frac{2\Delta_{4}\left(P_{2}+\beta_{1}\right)\left(A_{2}+A_{4}\right)}{\left(E_{c}-\Delta_{1}+\Delta_{2}\right)^{2}} \, .
\end{eqnarray}

This analytical approach for the conduction band provides a reasonable description up 
to 0.2 nm$^\textrm{-1}$ for InAs and 0.6 nm$^\textrm{-1}$ for InP, which is roughly 
100 meV above the energy gap in both cases. The numerical values of $m_z$, $m_{xy}$, 
$\alpha$, $\gamma_z$ and $\gamma_{xy}$ can be obtained by replacing the parameters with 
values presented in table~\ref{tab:par_full}. Setting the $k$-dependent SOC parameters  
$\gamma_1$ and $\beta_1$ to zero, we recover the analytical linear splitting found 
in Ref.~\onlinecite{Fu2008:JAP}. Our approach has the advantage of also providing 
the analytical description of the cubic terms. For additional corrections to the 
cubic term, it is possible to include higher order terms in the folding down approach. 
The comparison to {\it ab initio} data using the analytical expressions presented 
in this section can be found in the supplemental material.

%===============================================================================

\section{Compact description for valence band}
\label{sec:compactVB}

Because of the coupling from the crystal field and SOC energies, the best simplified 
description for the valence band is simply neglecting the coupling with conduction 
band, thus leading to a 6$\times$6 matrix. It is possible to write this 6$\times$6 
Hamiltonian in a compact form using direct products of 3$\times$3 (orbital) and 
2$\times$2 (spin) matrices.\cite{Chuang1996:PRB,Punya2012:PRB} In the basis set
$\left\{\left|c_{1}\right\rangle, \left|c_{3}\right\rangle, \left|c_{2}\right\rangle, \left|c_{5}\right\rangle, \left|c_{6}\right\rangle, \left|c_{4}\right\rangle \right\}$, 
the compact form of valence band is written as
\begin{eqnarray}
H_{\textrm{VB}} & \!\!\!=\!\!\! & \Delta_{1}J_{z}^{2}\mathcal{I}_{2}+\Delta_{2}J_{z}\sigma_{z}+\sqrt{2}\Delta_{3}\!\left(J_{+}\sigma_{-}+J_{-}\sigma_{+}\right)\nonumber \\
 & \!\!\!+\!\!\! & \left(A_{1}\mathcal{I}_{3}+A_{3}J_{z}^{2}\right)\!k_{z}^{2}\mathcal{I}_{2}+\left(A_{2}\mathcal{I}_{3}+A_{4}J_{z}^{2}\right)\!\!\left(k_{x}^{2}+k_{y}^{2}\right)\!\mathcal{I}_{2}\nonumber \\
 & \!\!\!-\!\!\! & A_{5}\!\left(J_{+}^{2}k_{-}^{2}+J_{-}^{2}k_{+}^{2}\right)\!\mathcal{I}_{2}\nonumber \\
 & \!\!\!-\!\!\! & 2A_{6}k_{z}\!\left(\left\{ J_{z}J_{+}\right\} k_{-}+\left\{ J_{z},J_{-}\right\} k_{+}\right)\mathcal{I}_{2\!}\nonumber \\
 & \!\!\!+\!\!\! & iA_{7}\left(J_{+}k_{-}-J_{-}k_{+}\right)\mathcal{I}_{2}\nonumber \\
 & \!\!\!+\!\!\! & i\sqrt{2}\alpha_{1}\left[\left\{ J_{z}J_{-}\right\} \!\left(\sigma_{z}k_{+}-2\sigma_{+}k_{z}\right)\right.\nonumber \\
 &               & \,\quad\quad\left.-\left\{ J_{z}J_{+}\right\} \!\left(\sigma_{z}k_{-}-2\sigma_{-}k_{z}\right)\right]\nonumber \\
 & \!\!\!+\!\!\! & i\left[\left(\alpha_{3}-\alpha_{2}\right)J_{z}^{2}-\alpha_{3}\mathcal{I}_{3}\right]\!\left(\sigma_{+}k_{-}-\sigma_{-}k_{+}\right)
\end{eqnarray}
with $\left\{J_aJ_b\right\}=\frac{1}{2}\left(J_aJ_b+J_bJ_a\right)$, $J_{\pm}=\frac{1}{\sqrt{2}}\left(J_{x}\pm J_{y}\right)$, 
$\sigma_{\pm}=\frac{1}{2}\left(\sigma_{x}\pm\sigma_{y}\right)$ and $k_\pm = k_x \pm i k_y$. 
The definitions of $J_x,J_y$ and $J_z$ matrices can be found in appendix A (eq. A3) 
of Ref.~\onlinecite{Chuang1996:PRB}. The matrix $\mathcal{I}_n$ is a $n$-dimensional 
identity. The product of 3$\times$3 matrices ($A$) with 2$\times$2 matrices ($a$) 
is defined here as
\begin{equation}
Aa=\left[\begin{array}{cc}
a_{11}A & a_{12}A\\
a_{22}A & a_{22}A
\end{array}\right] \, .
\end{equation}

\begin{table}[h!]
\caption{Parameter sets of the 6$\times$6 valence band Hamiltonian for InAs and 
InP WZ. The units follow Table~\ref{tab:par_full}.}
\begin{center}
{\renewcommand{\arraystretch}{1.2}
\begin{tabular*}{1\columnwidth}{@{\extracolsep{\fill}}ccc}
\hline
\hline
Parameter & InAs & InP \tabularnewline
\hline
Energy splittings &         &         \tabularnewline
$\Delta_{1}$      & 0.1003  & 0.0945  \tabularnewline
$\Delta_{2}$      & 0.1038  & 0.0286  \tabularnewline
$\Delta_{3}$      & 0.1037  & 0.0310  \tabularnewline
\hline
Linear parameters &         &         \tabularnewline
$A_{7}$           & -0.5565 & -0.0917 \tabularnewline
$\alpha_{1}$      & -0.0237 &  0.3309 \tabularnewline
$\alpha_{2}$      & -0.0758 & -0.0702 \tabularnewline
$\alpha_{3}$      & -0.0967 & -0.0521 \tabularnewline
\hline
Second order parameters &         &     \tabularnewline
$A_{1}$           & -17.2689 & -10.5414 \tabularnewline
$A_{2}$           &  -1.2047 &  -1.4542 \tabularnewline
$A_{3}$           &  16.6637 &   9.4589 \tabularnewline
$A_{4}$           &  -7.6202 &  -3.2741 \tabularnewline
$A_{5}$           &  -5.9281 &   3.9468 \tabularnewline
$A_{6}$           &  -7.3872 &  -0.2759 \tabularnewline
\hline
\hline
\end{tabular*}}
\end{center}
\label{tab:par_VB}
\end{table}

To obtain the best parameter sets that describe the {\it ab initio} band structure, 
we performed the same fitting approach described in section~\ref{sec:fit8}. We found 
that, in order to attain the monotonic behavior of the bands, some features of the 
band structure or the spin splittings are not matched as precisely as the results 
using the 8$\times$8 Hamiltonian. For instance, the band structures and the spin 
splittings for InP looks reasonable, however, the spin orientation for LH and CH 
shows opposite trends. For InAs, the spin texture follows the correct behavior, 
however, the band structure and the spin splittings show the {\it ab initio} features 
shifted to higher $k$ values. We show the fitting results for the 6$\times$6 description 
and the comparison to {\it ab initio} in the supplemental material. The best parameter 
sets are displayed in table~\ref{tab:par_VB}. We would like to emphasize that the 
most reliable approach is to use the 8$\times$8 Hamiltonian with parameter sets 
we provide in section \ref{sec:fit8}.

%===============================================================================

\section{Conclusions}
\label{sec:conclude}

In this paper, we have calculated the band structure of InAs and InP in WZ phase 
using the WIEN2k {\it ab initio} code. Both compounds have a direct band gap at $\Gamma$ 
point with the SOC effects clearly larger for InAs than InP. Our calculations are 
consistent with theoretical and experimental reported values in the literature. 
In order to describe the band structure around the FBZ center, we developed a multiband 
8$\times$8 {\bf k.p} model for the first conduction band and the top three valence 
bands, including spin. The fitted parameters we obtained for the {\bf k.p} Hamiltonian 
recover the important features of the {\it ab initio} band structure with good 
agreement up to 1.0 nm$^\textrm{-1}$ for multiple directions in the FBZ. Due to 
the stronger SOC of InAs compared to its crystal field splitting, the labeling of 
LH and CH energy bands at $\Gamma$ point is reversed from InP. Regarding the spin 
splitting properties, we included the $k$-dependent SOC term in the Hamiltonian, 
which is usually neglected in the literature. This term, combined with the other 
indirect couplings in the Hamiltonian, allowed the description of the spin splitting 
properties further away from the vicinity of $\Gamma$ point. Our model captured 
all the important features including the description of maxima values and also 
the crossing between the spin split bands (clearly seen in HH band of InP, for instance). 
All these intricate behaviors of spin splitting have a unique physical origin, the 
BIA of the WZ structure. Furthermore, we calculated the in-plane spin orientation, 
i.~e., the spin expectation value, of the energy bands and found that they all have 
a Rashba-like spin texture, either CW or CCW. This spin orientation was also compared 
to {\it ab initio} data to correctly identify the signs of the parameters in the 
Hamiltonian. Using our multiband {\bf k.p} Hamiltonian, we obtained the DOS for 
conduction and valence bands and calculated the carrier density as function of the 
Fermi energy. In addition to the 8$\times$8 Hamiltonian, we present analytical expressions 
for the effective masses and the SOC field of conduction band which holds in the 
vicinities of the $\Gamma$ point. For completeness, we also fitted the 6$\times$6 
{\bf k.p} model for valence band to the {\it ab initio} data. We emphasize that 
the best effective description that matches our {\it ab initio} calculations 
is the full 8$\times$8 {\bf k.p} Hamiltonian.

In conclusion, we provided in this study robust {\bf k.p} models and parameter sets 
that can be straightforwardly applied to investigate novel effects in InAs- and 
InP-based nanostructures. For instance, polytypic systems of mixed WZ and ZB are already 
demonstrated experimentally for both InAs and InP with great growth control of the 
different phases\cite{Caroff2009:NatNano} and there are also theoretical models 
to treat such systems.\cite{FariaJunior2012:JAP,FariaJunior2014:JAP,Climente2016:JAP} 
Furthermore, InAs nanowires are also a platform for studies in Majorana fermions\cite{Das2012:NatPhys}. 
One of the key ingredients for such realization is the presence of a robust SOC 
to split the energy bands, a feature already included in our model. Finally, it 
is straightforward to included strain effects by using the well established WZ strain 
Hamiltonian\cite{Chuang1996:PRB,FariaJunior2012:JAP} combined with the elastic constants 
and deformation potentials for InAs and InP in WZ phase already reported in the literature.
\cite{Larsson2007:Nanotech,Boxberg2012:AdvMat,Hajlaoui2013:JPdAP,Hajlaoui2015:JETP}

%===============================================================================

\section*{Acknowledgements}

The authors acknowledge financial support to CAPES PVE (Grant No. 88881.068174/2014-01), 
CNPq (Grants No. 149904/2013-4, 88887.110814/2015-00 and 304289/2015-9), DFG SFB 
689 and FAPESP (Grant No. 2012/05618-0). PEFJ thanks A. Polimeni for suggesting 
the calculation of effective masses.

%===============================================================================

\section*{Appendix A: effective masses}

Very close to $\Gamma$-point we can estimate the effective 
masses by fitting a parabolic dispersion to the {\it ab initio} data. In table \ref{tab:effective_masses}, 
we show the values of effective masses along $k_z$ and $k_x$ directions for the 
highlighted energy bands of Figs.~\ref{fig:zoom_bs}(a)-(b). For $k_x$ direction, 
we calculated the effective masses assuming the average value of the spin splitting 
bands, i. e., $\left(E_o + E_i\right)/2$ with the subindex $o\;(i)$ indicating the 
outer (inner) branch.

\begin{table}[h!]
\caption{Effective masses for InAs and InP along $k_z$ ($m_\parallel^*$) and $k_x$ 
($m_\perp^*$) for the highlighted bands of Fig. \ref{fig:zoom_bs}(a)-(b). The effective 
masses were obtained by fitting a parabola up to 2\% of the FBZ along the specified 
directions.}
\begin{center}
{\renewcommand{\arraystretch}{1.2}
\begin{tabular*}{1\columnwidth}{@{\extracolsep{\fill}}lcccc}
\hline 
\hline
 & \multicolumn{2}{c}{InAs} & \multicolumn{2}{c}{InP}\tabularnewline
 & $m_{\parallel}^{*}$ & $m_{\perp}^{*}$ & $m_{\parallel}^{*}$ & $m_{\perp}^{*}$\tabularnewline
\hline 
$\Gamma_{7c}$          &  0.0370 &  0.0416 &  0.0947 &  0.1183\tabularnewline
$\Gamma_{9v}$          & -0.9738 & -0.0795 & -1.0646 & -0.2091\tabularnewline
$\Gamma_{7v}$          & -0.0551 & -0.1046 & -0.3064 & -0.1988\tabularnewline
$\Gamma_{7v}^{\prime}$ & -0.0863 & -0.1838 & -0.1016 & -0.4887\tabularnewline
\hline
\hline
\end{tabular*}}
\end{center}
\label{tab:effective_masses}
\end{table}

%===============================================================================

\section*{Appendix B: Hamiltonian terms and parameters}

In this appendix, we present the matrix forms of all terms in equation (\ref{eq:Hmatrix}) 
and the definition of parameters using the simple group formalism.

Matrix representation of $H_0$:
\begin{equation}
H_{0} = \textrm{diag} \left[ \Delta_1, \, \Delta_1, \, 0, \, \Delta_1, \, \Delta_1, \, 0, E_c, \, E_c \right] \, ,
\end{equation}
with the definitions $\left\langle \Gamma_{5v}^x \left|H_{0}\right| \Gamma_{5v}^x \right\rangle =\left\langle \Gamma_{5v}^y \left|H_{0}\right|\Gamma_{5v}^y\right\rangle =\Delta_{1}$, 
$\left\langle \Gamma_{1v}\left|H_{0}\right|\Gamma_{1v}\right\rangle =0$ and $\left\langle \Gamma_{1c}\left|H_{0}\right|\Gamma_{1c}\right\rangle =E_c$.

The zero energy is defined without SOC for states $\left|c_{3}\right\rangle$ and 
$\left|c_{6}\right\rangle$. The parameter $\Delta_1$ is the crystal field splitting 
energy, which arises due to the WZ anisotropy between $xy$ plane and $z$ direction, 
and the conduction band energy is denoted by the parameter $E_c$. It is possible 
to make the connection with the energy gap including SOC coupling by writing 
$E_c = E_g + \Delta_c$, for instance. It is also convenient to consider a diagonal 
energy offset to set the top valence band at zero energy.

\begin{widetext}
Matrix representation of $H^{(1)}_{SO}$:
\begin{equation}
H_{SO}^{(1)}=\left[\begin{array}{cccccccc}
\Delta_{2} & 0 & 0 & 0 & 0 & 0 & 0 & 0\\
0 & -\Delta_{2} & 0 & 0 & 0 & \sqrt{2}\Delta_{3} & 0 & i\sqrt{2}\Delta_{4}\\
0 & 0 & 0 & 0 & \sqrt{2}\Delta_{3} & 0 & 0 & 0\\
0 & 0 & 0 & \Delta_{2} & 0 & 0 & 0 & 0\\
0 & 0 & \sqrt{2}\Delta_{3} & 0 & -\Delta_{2} & 0 & i\sqrt{2}\Delta_{4} & 0\\
0 & \sqrt{2}\Delta_{3} & 0 & 0 & 0 & 0 & 0 & 0\\
0 & 0 & 0 & 0 & -i\sqrt{2}\Delta_{4} & 0 & 0 & 0\\
0 & -i\sqrt{2}\Delta_{4} & 0 & 0 & 0 & 0 & 0 & 0
\end{array}\right] \, ,
\end{equation}
with the definitions
\begin{eqnarray}
\Delta_{2} & = & \frac{i\hbar}{4m_{0}^{2}c^{2}}\left\langle \Gamma_{5v}^x \left|\frac{\partial V}{\partial x}p_{y}-\frac{\partial V}{\partial y}p_{x}\right| \Gamma_{5v}^y \right\rangle \nonumber \\
\Delta_{3} & = & \frac{i\hbar}{4m_{0}^{2}c^{2}}\left\langle \Gamma_{5v}^y \left|\frac{\partial V}{\partial y}p_{z}-\frac{\partial V}{\partial z}p_{y}\right| \Gamma_{1v}   \right\rangle = \frac{i\hbar}{4m_{0}^{2}c^{2}}\left\langle \Gamma_{1v} \left|\frac{\partial V}{\partial z}p_{x}-\frac{\partial V}{\partial x}p_{z}\right| \Gamma_{5v}^x \right\rangle \nonumber \\
\Delta_{4} & = & \frac{i\hbar}{4m_{0}^{2}c^{2}}\left\langle \Gamma_{5v}^y \left|\frac{\partial V}{\partial y}p_{z}-\frac{\partial V}{\partial z}p_{y}\right| \Gamma_{1c}   \right\rangle = \frac{i\hbar}{4m_{0}^{2}c^{2}}\left\langle \Gamma_{1c} \left|\frac{\partial V}{\partial z}p_{x}-\frac{\partial V}{\partial x}p_{z}\right| \Gamma_{5v}^x \right\rangle \, .
\end{eqnarray}

Matrix representation of $H^{(1)}_{kp}$:
\begin{equation}
H_{kp}^{(1)}=\left[\begin{array}{cccccccc}
0 & 0 & iA_{7}k_{-} & 0 & 0 & 0 & -\frac{1}{\sqrt{2}}P_{2}k_{-} & 0\\
0 & 0 & -iA_{7}k_{+} & 0 & 0 & 0 & \frac{1}{\sqrt{2}}P_{2}k_{+} & 0\\
-iA_{7}k_{+} & iA_{7}k_{-} & 0 & 0 & 0 & 0 & P_{1}k_{z} & 0\\
0 & 0 & 0 & 0 & 0 & -iA_{7}k_{+} & 0 & \frac{1}{\sqrt{2}}P_{2}k_{+}\\
0 & 0 & 0 & 0 & 0 & iA_{7}k_{-} & 0 & -\frac{1}{\sqrt{2}}P_{2}k_{-}\\
0 & 0 & 0 & iA_{7}k_{-} & -iA_{7}k_{+} & 0 & 0 & P_{1}k_{z}\\
-\frac{1}{\sqrt{2}}P_{2}k_{+} & \frac{1}{\sqrt{2}}P_{2}k_{-} & P_{1}k_{z} & 0 & 0 & 0 & 0 & 0\\
0 & 0 & 0 & \frac{1}{\sqrt{2}}P_{2}k_{-} & -\frac{1}{\sqrt{2}}P_{2}k_{+} & P_{1}k_{z} & 0 & 0
\end{array}\right] \, ,
\end{equation}
with the definitions
\begin{eqnarray}
A_{7} & = & \frac{i}{\sqrt{2}}\frac{\hbar}{m_{0}}\left\langle \Gamma_{5v}^x \left|p_{x}\right|\Gamma_{1v}\right\rangle =\frac{i}{\sqrt{2}}\frac{\hbar}{m_{0}}\left\langle \Gamma_{5v}^y \left|p_{y}\right|\Gamma_{1v}\right\rangle \nonumber \\
P_{2} & = & i\frac{\hbar}{m_{0}}\left\langle \Gamma_{5v}^x \left|p_{x}\right|\Gamma_{1c}\right\rangle =i\frac{\hbar}{m_{0}}\left\langle \Gamma_{5v}^y\left|p_{y}\right|\Gamma_{1c}\right\rangle \nonumber \\
P_{1} & = & i\frac{\hbar}{m_{0}}\left\langle \Gamma_{1v}\left|p_{z}\right|\Gamma_{1c}\right\rangle \nonumber \\
k_{\pm} & = & k_{x}\pm ik_{y} \, .
\end{eqnarray}

Matrix representation of $H^{(1)}_{kSO}$:
\begin{equation}
H_{kSO}^{(1)}=\left[\begin{array}{cccccccc}
0 & 0 & -\frac{i}{\sqrt{2}}\alpha_{1}k_{-} & 0 & -i\alpha_{2}k_{-} & 0 & \frac{1}{\sqrt{2}}\beta_{1}k_{-} & 0\\
0 & 0 & -\frac{i}{\sqrt{2}}\alpha_{1}k_{+} & -i\alpha_{2}k_{-} & 0 & i\sqrt{2}\alpha_{1}k_{z} & \frac{1}{\sqrt{2}}\beta_{1}k_{+} & -\sqrt{2}\beta_{1}k_{z}\\
\frac{i}{\sqrt{2}}\alpha_{1}k_{+} & \frac{i}{\sqrt{2}}\alpha_{1}k_{-} & 0 & 0 & -i\sqrt{2}\alpha_{1}k_{z} & -i\alpha_{3}k_{-} & 0 & \beta_{2}k_{-}\\
0 & i\alpha_{2}k_{+} & 0 & 0 & 0 & \frac{i}{\sqrt{2}}\alpha_{1}k_{+} & 0 & -\frac{1}{\sqrt{2}}\beta_{1}k_{+}\\
i\alpha_{2}k_{+} & 0 & i\sqrt{2}\alpha_{1}k_{z} & 0 & 0 & \frac{i}{\sqrt{2}}\alpha_{1}k_{-} & -\sqrt{2}\beta_{1}k_{z} & -\frac{1}{\sqrt{2}}\beta_{1}k_{-}\\
0 & -i\sqrt{2}\alpha_{1}k_{z} & i\alpha_{3}k_{+} & -\frac{i}{\sqrt{2}}\alpha_{1}k_{-} & -\frac{i}{\sqrt{2}}\alpha_{1}k_{+} & 0 & -\beta_{2}k_{+} & 0\\
\frac{1}{\sqrt{2}}\beta_{1}k_{+} & \frac{1}{\sqrt{2}}\beta_{1}k_{-} & 0 & 0 & -\sqrt{2}\beta_{1}k_{z} & -\beta_{2}k_{-} & 0 & -i\gamma_{1}k_{-}\\
0 & -\sqrt{2}\beta_{1}k_{z} & \beta_{2}k_{+} & -\frac{1}{\sqrt{2}}\beta_{1}k_{-} & -\frac{1}{\sqrt{2}}\beta_{1}k_{+} & 0 & i\gamma_{1}k_{+} & 0
\end{array}\right] \, ,
\end{equation}
\end{widetext}
with the definitions
\begin{eqnarray}
\alpha_{1} & = & \frac{\hbar^{2}}{4m_{0}^{2}c^{2}}\left\langle \Gamma_{5v}^x \left|\frac{\partial V}{\partial x}\right| \Gamma_{1v}   \right\rangle =\frac{\hbar^{2}}{4m_{0}^{2}c^{2}}\left\langle \Gamma_{5v}^y \left|\frac{\partial V}{\partial y}\right| \Gamma_{1v}   \right\rangle \nonumber \\
\alpha_{2} & = & \frac{\hbar^{2}}{4m_{0}^{2}c^{2}}\left\langle \Gamma_{5v}^x \left|\frac{\partial V}{\partial z}\right| \Gamma_{5v}^x \right\rangle =\frac{\hbar^{2}}{4m_{0}^{2}c^{2}}\left\langle \Gamma_{5v}^y \left|\frac{\partial V}{\partial z}\right| \Gamma_{5v}^y \right\rangle  \nonumber \\
\alpha_{3} & = & \frac{\hbar^{2}}{4m_{0}^{2}c^{2}}\left\langle \Gamma_{1v}   \left|\frac{\partial V}{\partial z}\right| \Gamma_{1v}   \right\rangle  \nonumber \\
\beta_{1}  & = & \frac{\hbar^{2}}{4m_{0}^{2}c^{2}}\left\langle \Gamma_{5v}^x \left|\frac{\partial V}{\partial x}\right| \Gamma_{1c}   \right\rangle =\frac{\hbar^{2}}{4m_{0}^{2}c^{2}}\left\langle \Gamma_{5v}^y \left|\frac{\partial V}{\partial y}\right| \Gamma_{1c}   \right\rangle  \nonumber \\
\beta_{2}  & = & \frac{\hbar^{2}}{4m_{0}^{2}c^{2}}\left\langle \Gamma_{1v}   \left|\frac{\partial V}{\partial z}\right| \Gamma_{1c}   \right\rangle  \nonumber \\
\gamma_{1} & = & \frac{\hbar^{2}}{4m_{0}^{2}c^{2}}\left\langle \Gamma_{1c}   \left|\frac{\partial V}{\partial z}\right| \Gamma_{1c}   \right\rangle \, .
\end{eqnarray}

Matrix representation of $H^{(2)}_{kp}$:
\begin{equation}
H_{kp}^{(2)}=\left[\begin{array}{cccccccc}
\lambda+\theta & -K^{*} & -H^{*} & 0 & 0 & 0 & T^{*} & 0\\
-K & \lambda+\theta & H & 0 & 0 & 0 & T & 0\\
-H & H^{*} & \lambda & 0 & 0 & 0 & U & 0\\
0 & 0 & 0 & \lambda+\theta & -K & H & 0 & T\\
0 & 0 & 0 & -K^{*} & \lambda+\theta & -H^{*} & 0 & T^{*}\\
0 & 0 & 0 & H^{*} & -H & \lambda & 0 & U\\
T & T^{*} & U^{*} & 0 & 0 & 0 & V & 0\\
0 & 0 & 0 & T^{*} & T & U^{*} & 0 & V
\end{array}\right] \, ,
\end{equation}
with elements given by
\begin{eqnarray}
\lambda & = & A_{1}k_{z}^{2}+A_{2}\left(k_{x}^{2}+k_{y}^{2}\right)\nonumber \\
\theta & = & A_{3}k_{z}^{2}+A_{4}\left(k_{x}^{2}+k_{y}^{2}\right)\nonumber \\
K & = & A_{5}k_{+}^{2}\nonumber \\
H & = & A_{6}k_{+}k_{z}\nonumber \\
T & = & iB_{3}k_{+}k_{z}\nonumber \\
U & = & i\left[B_{1}k_{z}^{2}+B_{2}\left(k_{x}^{2}+k_{y}^{2}\right)\right]\nonumber \\
V & = & e_{1}k_{z}^{2}+e_{2}\left(k_{x}^{2}+k_{y}^{2}\right)
\end{eqnarray}
and all the parameters in units of $\hbar^2/2m_0$.

The term $\mathbf{H_{k2}}$ is already included in the diagonal terms of $H_{kp}^{(2)}$. 
Strictly speaking, the matrix representation $H_{kp}^{(2)}$ is defined as
\begin{equation}
H_{kp,\alpha\alpha'}^{(2)} = \mathbf{H_{k2}}\delta_{\alpha\alpha^\prime} + \sum_{\beta}^{B}\frac{\left\langle \alpha\left| \mathbf{H_{kp}} \right|\beta\right\rangle \left\langle \beta\left| \mathbf{H_{kp}} \right|\alpha'\right\rangle }{E_{\alpha\alpha'}-E_{\beta}} \, .
\end{equation}

The second order {\bf k.p} parameters $A_1$ to $A_6$ are defined the same way as Ref.~\onlinecite{Chuang1996:PRB}, 
while $e_1$, $e_2$, $B_1$, $B_2$ and $B_3$ are given by
\begin{eqnarray}
e_{1} & = & 1+\frac{2}{m_{0}}\sum_{\beta}^{B\left[\Gamma_{1}\right]}\frac{\left|\left\langle \Gamma_{1c}\left|p_{z}\right|\beta\right\rangle \right|^{2}}{E_{1c}-E_{\beta}}\nonumber \\
e_{2} & = & 1+\frac{2}{m_{0}}\sum_{\beta}^{B\left[\Gamma_{5}\right]}\frac{\left|\left\langle \Gamma_{1c}\left|p_{x}\right|\beta\right\rangle \right|^{2}}{E_{1c}-E_{\beta}}\nonumber \\
 & = & 1+\frac{2}{m_{0}}\sum_{\beta}^{B\left[\Gamma_{5}\right]}\frac{\left|\left\langle \Gamma_{1c}\left|p_{y}\right|\beta\right\rangle \right|^{2}}{E_{1c}-E_{\beta}}\nonumber \\
B_{1} & = & \frac{2}{m_{0}}\sum_{\beta}^{B\left[\Gamma_{1}\right]}\frac{\left\langle \Gamma_{1v}\left|p_{z}\right|\beta\right\rangle \left\langle \beta\left|p_{z}\right|\Gamma_{1c}\right\rangle }{E_{1v1c}-E_{\beta}}\nonumber \\
B_{2} & = & \frac{2}{m_{0}}\sum_{\beta}^{B\left[\Gamma_{5}\right]}\frac{\left\langle \Gamma_{1v}\left|p_{x}\right|\beta\right\rangle \left\langle \beta\left|p_{x}\right|\Gamma_{1c}\right\rangle }{E_{1v1c}-E_{\beta}}\nonumber \\
 & = & \frac{2}{m_{0}}\sum_{\beta}^{B\left[\Gamma_{5}\right]}\frac{\left\langle \Gamma_{1v}\left|p_{y}\right|\beta\right\rangle \left\langle \beta\left|p_{y}\right|\Gamma_{1c}\right\rangle }{E_{1v1c}-E_{\beta}}\nonumber \\
B_{3} & = & \frac{\sqrt{2}}{m_{0}}\left(\sum_{\beta}^{B\left[\Gamma_{1}\right]}\frac{\left\langle \Gamma_{5v}^{x}\left|p_{x}\right|\beta\right\rangle \left\langle \beta\left|p_{z}\right|\Gamma_{1c}\right\rangle }{E_{5v1c}-E_{\beta}}\right.\nonumber \\
 &  & \quad\;\;+\left.\sum_{\beta}^{B\left[\Gamma_{5}\right]}\frac{\left\langle \Gamma_{5v}^{x}\left|p_{z}\right|\beta\right\rangle \left\langle \beta\left|p_{x}\right|\Gamma_{1c}\right\rangle }{E_{5v1c}-E_{\beta}}\right)\nonumber \\
 & = & \frac{\sqrt{2}}{m_{0}}\left(\sum_{\beta}^{B\left[\Gamma_{1}\right]}\frac{\left\langle \Gamma_{5v}^{y}\left|p_{y}\right|\beta\right\rangle \left\langle \beta\left|p_{z}\right|\Gamma_{1c}\right\rangle }{E_{5v1c}-E_{\beta}}\right.\nonumber \\
 &  & \quad\;\;+\left.\sum_{\beta}^{B\left[\Gamma_{5}\right]}\frac{\left\langle \Gamma_{5v}^{y}\left|p_{z}\right|\beta\right\rangle \left\langle \beta\left|p_{y}\right|\Gamma_{1c}\right\rangle }{E_{5v1c}-E_{\beta}}\right) \, ,
\end{eqnarray} 
with non-zero contributions represented by the irreducible representations in the 
brackets above the summation.

%===============================================================================

\section*{Appendix C: fitting in other directions}

\begin{figure}[h!]
\begin{center} 
\includegraphics{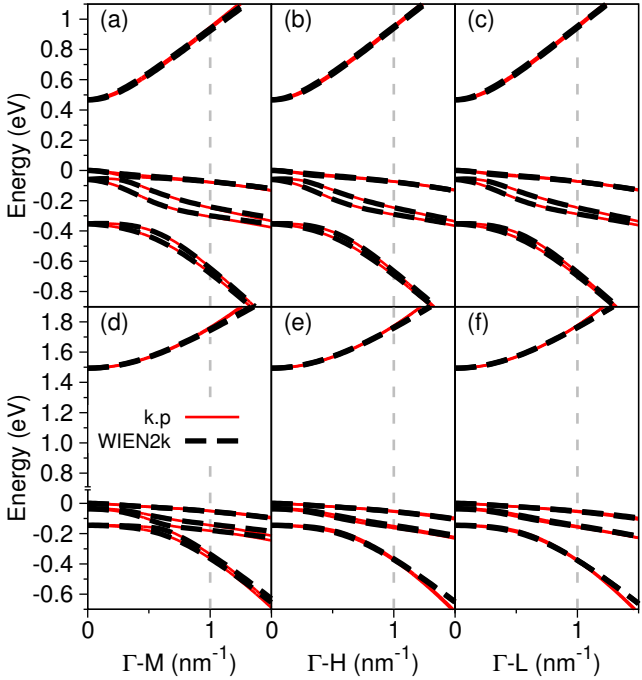}
\caption{(Color online) Comparison of band structures for (a) $\Gamma$-M, (b) $\Gamma$-H and (c) 
$\Gamma$-L directions of InAs and (d) $\Gamma$-M, (d) $\Gamma$-H and (f) $\Gamma$-L 
directions of InP. The line schemes follow Fig.~\ref{fig:fitted_bs}.}
\label{fig:append_fitted_bs}
\end{center}
\end{figure}

The comparison between the fitted and WIEN2k {\it ab initio} band structures is displayed 
in Fig.~\ref{fig:append_fitted_bs} for $\Gamma$-M, $\Gamma$-H and $\Gamma$-L directions. 
For $\Gamma$-M direction, we have the same behavior discussed for $\Gamma$-K. However, 
the {\bf k.p} band structure in $\Gamma$-H and $\Gamma$-L directions have closer 
values to {\it ab initio}. This better agreement arises from the second order parameters 
$A_6$ and $B_3$ which only couple $k_x k_y$ plane with $k_z$, providing additional 
corrections to the band structures.

\begin{figure}[h!]
\begin{center} 
\includegraphics{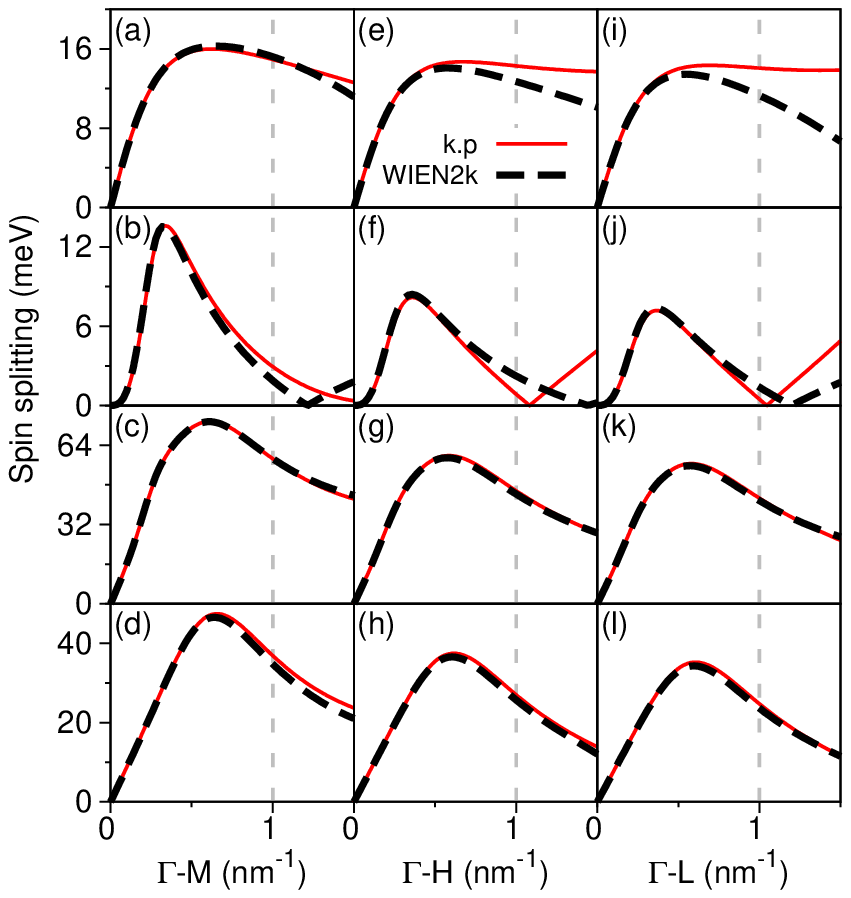}
\caption{(Color online) Comparison of the InAs spin splittings for (a,e,i) CB, (b,f,j) 
HH, (c,g,k) CH and (d,h,l) LH along $\Gamma$-M, $\Gamma$-H and $\Gamma$-L direcions, 
respectively. The line schemes follow Fig.~\ref{fig:fitted_bs}.}
\label{fig:append_fitted_sp_InAs}
\end{center}
\end{figure}

\begin{figure}[h!]
\begin{center} 
\includegraphics{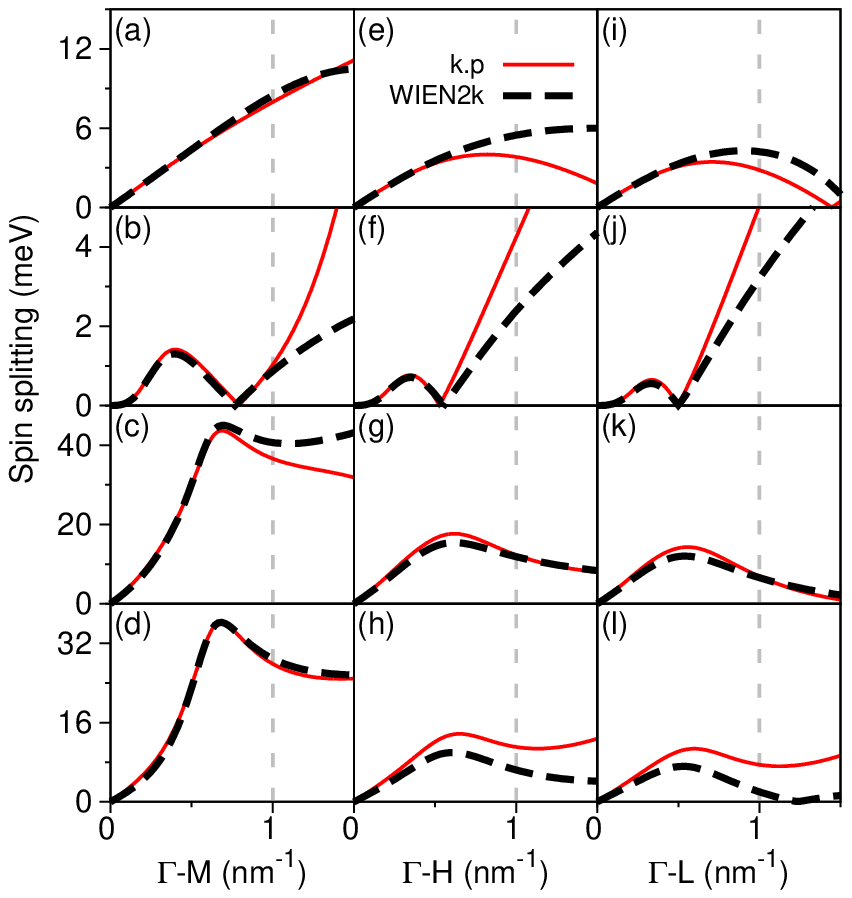}
\caption{(Color online) Comparison of the InP spin splittings for (a,e,i) CB, (b,f,j) 
HH, (c,g,k) LH and (d,h,l) CH along $\Gamma$-M, $\Gamma$-H and $\Gamma$-L directions, 
respectively. The line schemes follow Fig.~\ref{fig:fitted_bs}.}
\label{fig:append_fitted_sp_InP}
\end{center}
\end{figure}

In Fig.~\ref{fig:append_fitted_sp_InAs} and Fig.~\ref{fig:append_fitted_sp_InP}, 
we show the comparison of spin splittings along $\Gamma$-M, $\Gamma$-H and 
$\Gamma$-L directions for InAs and InP, respectively. For $\Gamma$-H and $\Gamma$-L
directions, the spin splittings are usually smaller compared to $\Gamma$-K and 
$\Gamma$-M. This difference, however, depends on the material and the energy band. 
For instance, CH and LH values for InAs in $\Gamma$-H and $\Gamma$-L are 
approximately half the value in $\Gamma$-K and $\Gamma$-M directions while LH 
and CH values for InP are approximately one fourth of the values. Because of 
this larger differences for InP, the spin splittings for CH bands along $\Gamma$-H 
and $\Gamma$-L show a small deviation compared to {\it ab initio} [Figs.~\ref{fig:append_fitted_sp_InP}(h) 
and \ref{fig:append_fitted_sp_InP}(l)]. However, the crossings for HH bands [Figs.~\ref{fig:append_fitted_sp_InP}(f) 
and \ref{fig:append_fitted_sp_InP}(j)] are precisely reproduced. Comparing all 
directions, we verify that our {\bf k.p} model and parameter sets reproduce with 
great agreement the {\it ab initio} band structure and spin splittings along all the 
considered directions of the FBZ.

%===============================================================================
%\bibitem{authorYear:journal} - papers
%\bibitem{authorYear}         - book
%\bibitem{note:something}     - footnotes

%===============================================================================

\clearpage

\onecolumngrid

\section*{\large{Supplemental material for the paper ``Realistic multiband {\bf k.p} approach from {\it ab initio} and spin-orbit coupling effects of InAs and InP in wurtzite phase''}}

\vspace{1cm}

%===============================================================================
\section*{\uppercase{I. Curve fitting for the carrier density}}

In this section, we provide a curve fitting of the carrier densities presented 
in Figs. 7(c) and 7(d) of the main paper. The 3D parabolic model gives us a carrier 
density dependence of the form $n(E) \propto E^{\frac{3}{2}}$, with the proportionality 
constant dependent on the effective mass. Since the calculated band structures of 
InAs and InP do not obey this parabolic behavior, we use the functional form
\begin{equation}
n(E) = a + bE + cE^d \, ,
\label{eq:fitn}
\end{equation}
to fit the carrier density as a function of the Fermi energy, with $E = E_f - E_g$ 
for the electrons and $E = \left|E_f\right|$ for the holes. The energy $E$ is given 
in meV and the carrier density $n$ in $10^{16}\;\textrm{cm}^{-3}$. This functional 
form does not carry any physical meaning in its different terms, it just provides 
an analytical way to predict the carrier density given the Fermi energy measured 
from the band edge. Because of linear SOC terms in conduction band, DOS is nonzero 
at the energy gap, therefore the requirement to use the fitting parameter $a$. Also, 
the linear dispersion $bE$ gives a better agreement with the numerical data. For 
valence band, the fitting of parameters $c$ and $d$ is enough, and therefore $a=b=0$. 
The best fitting is obtained by separating the carrier density in two different 
regions, which we call low and high energy regimes. The fitting parameters are shown 
in table~\ref{tab:fit}.

\begin{table}[H]
\caption{Numerical parameters to be used in equation (\ref{eq:fitn}) to predict 
the carrier density or the Fermi energy.}
\begin{center}
\begin{tabular}{lccccc|clcc}
\hline
\hline
\multicolumn{5}{c}{Electrons} &  &  & \multicolumn{3}{c}{Holes}\tabularnewline
\hline 
 & $a$ & $b$ & $c$ & $d$ &  &  &  & $c$ & $d$\tabularnewline
\hline 
InAs low ($E\leq35$) & 0.0385 & 0.1115 & 0.0684 & 1.6417 &  &  & InAs low ($E\leq6$) & 1.1076 & 1.5923\tabularnewline
InAs high ($35<E\leq300$) & -9.4227 & 0.8434 & 0.0023 & 2.2664 &  &  & InAs high ($6<E\leq100$) & 0.4065 & 2.1492\tabularnewline
\hline 
InP low ($E\leq35$) & 0.0117 & 0.1022 & 0.5636 & 1.5239 &  &  & InP low ($E\leq10$) & 2.4625 & 1.6694\tabularnewline
InP high ($35<E\leq300$) & -10.1466 & 1.5372 & 0.2265 & 1.6747 &  &  & InP high ($10<E\leq100$) & 0.9292 & 2.0845 \tabularnewline
\hline 
\end{tabular}
\end{center}
\label{tab:fit}
\end{table}

%===============================================================================

\section*{\uppercase{II. Analytical description for the conduction band compared to {\it ab initio}}}

\begin{figure}[h!]
\begin{center} 
\includegraphics{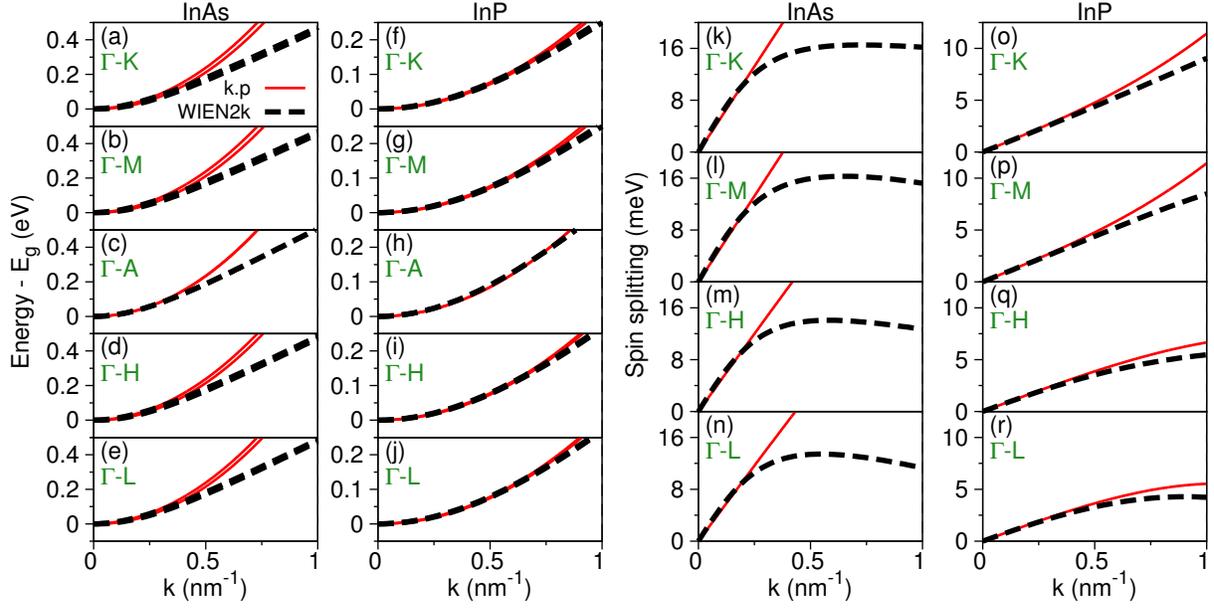}
\caption{(Color online) Comparison between the analytical expression for conduction 
band and WIEN2k band structure for (a-e) InAs and (f-j) InP for $\Gamma$-(K, M, A, H, L) 
directions. Comparison between the analytical spin splittings for (k-n) InAs and 
(o-r) InP along $\Gamma$-(K, M, H, L) directions.}
\label{fig:analyticalCB}
\end{center}
\end{figure}

Using the analytical expressions for the conduction band provided in the main paper 
(section V), we show the comparison between this approach and the {\it ab initio} 
data in Fig.~\ref{fig:analyticalCB}. Although the description for InP provides 
better results further away of $\Gamma$ point, the maximum accurate energy is 
around 100 meV above the gap for both materials.

%===============================================================================

\section*{\uppercase{III. 6$\times$6 model for the valence band compared to {\it ab initio}}}

\subsection*{A. InAs}

\begin{figure}[h!]
\begin{center} 
\includegraphics{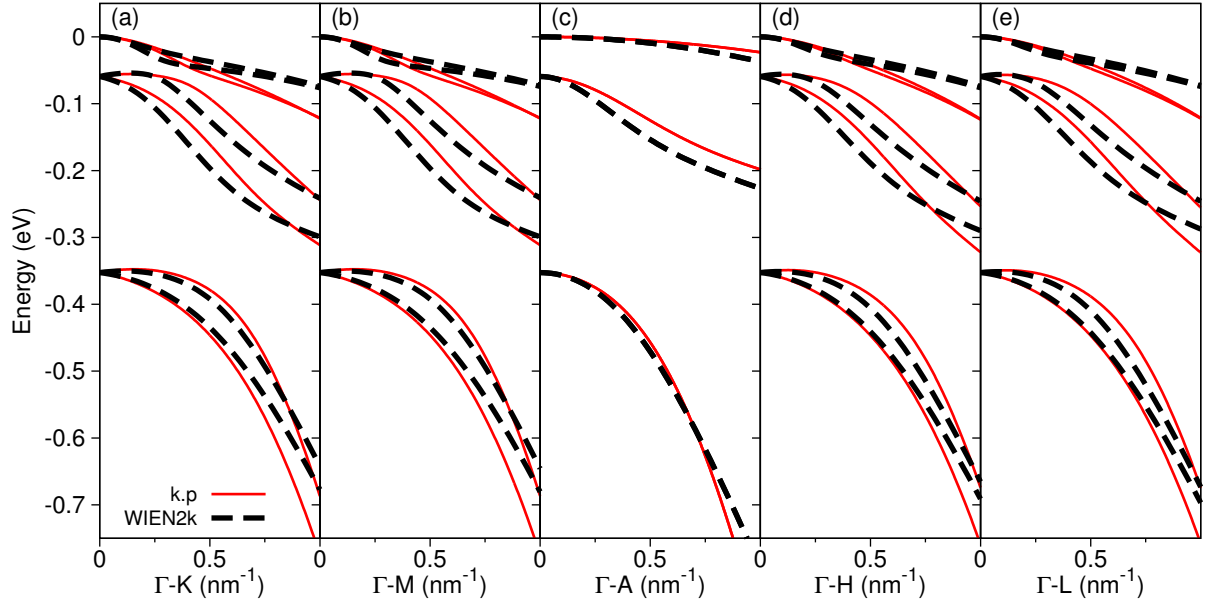}
\caption{(Color online) Comparison between the 6$\times$6 {\bf k.p} model for valence
band and WIEN2k band structure for InAs along (a) $\Gamma$-K, (b) $\Gamma$-M, (c) $\Gamma$-A, (d) 
$\Gamma$-H and (e) $\Gamma$-L.}
\label{fig:InAs_6x6_bs}
\end{center}
\end{figure}

\begin{figure}[h!]
\begin{center} 
\includegraphics{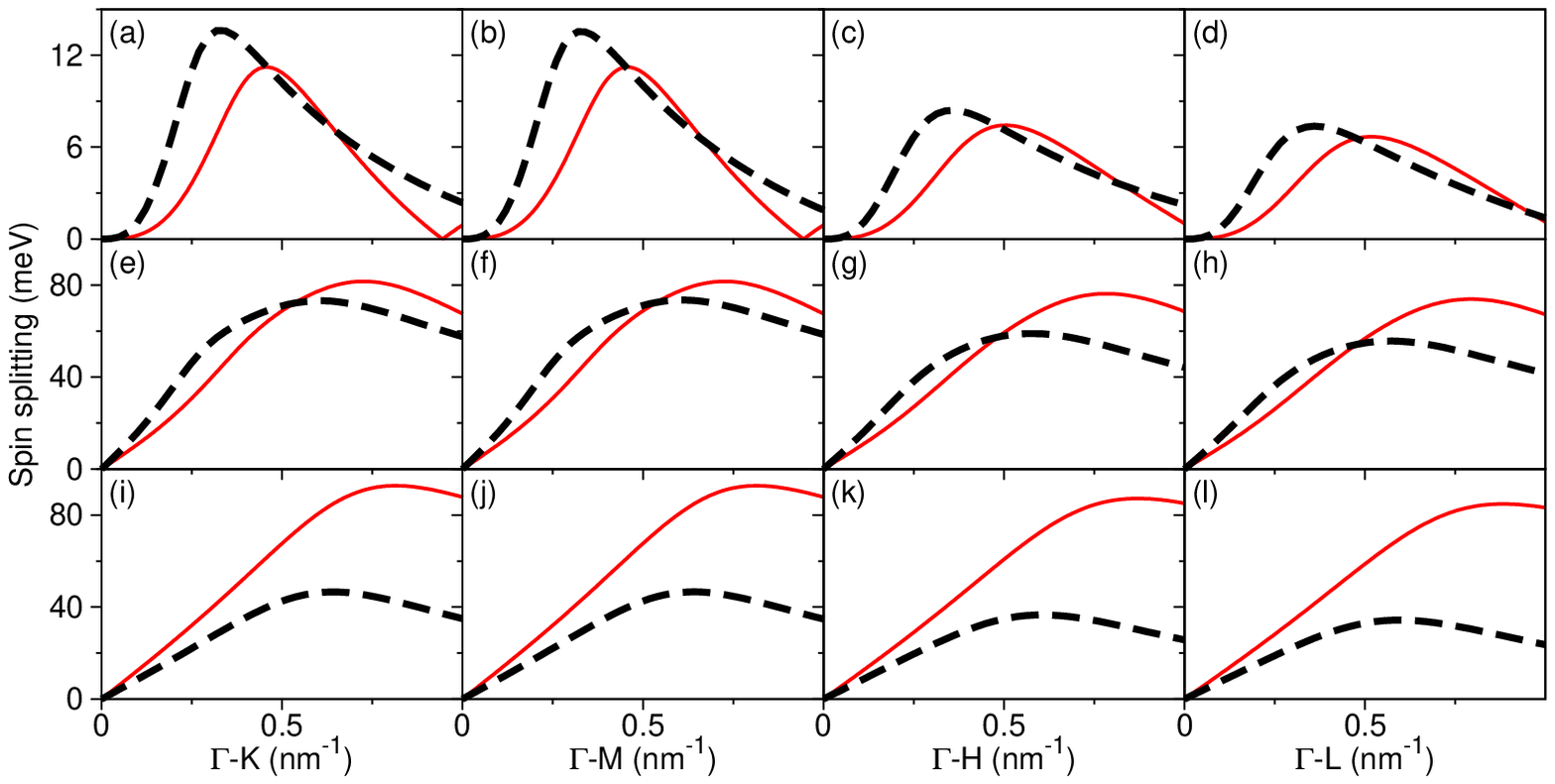}
\caption{(Color online) Comparison of the InAs spin splittings for (a-d) HH, 
(e-h) CH and (i-l) LH along $\Gamma$-(K, M, H, L) direcions.}
\label{fig:InAs_6x6_sp}
\end{center}
\end{figure}

Applying the fitting approach discussed in Sec. IV of the main paper for the valence 
band of InAs, we obtain the band structure displayed in Fig,~\ref{fig:InAs_6x6_bs}. 
In order to achieve the monotonic behavior of the valence band, i. e., decreasing 
energy while increasing $k$, the resulting fitted parameters provide a band structure 
that is shifted to higher $k$ values. These same trends can be seen in the spin 
splitting, shown in Fig,~\ref{fig:InAs_6x6_sp}, by looking at the peak values. 
The correct fitting closer to $\Gamma$ point does not provide the correct monotonic 
behavior, i. e., it either makes the spin splitting branches diverge drastically 
or the HH band acquires an upward curvature. Because of the small energy gap of 
InAs and the large SOC effects, we emphasize that including the explicit coupling 
with the conduction band is necessary and, therefore, the most suitable approach 
is the 8$\times$8 model we discuss in the main paper.

\begin{figure}[h!]
\begin{center} 
\includegraphics{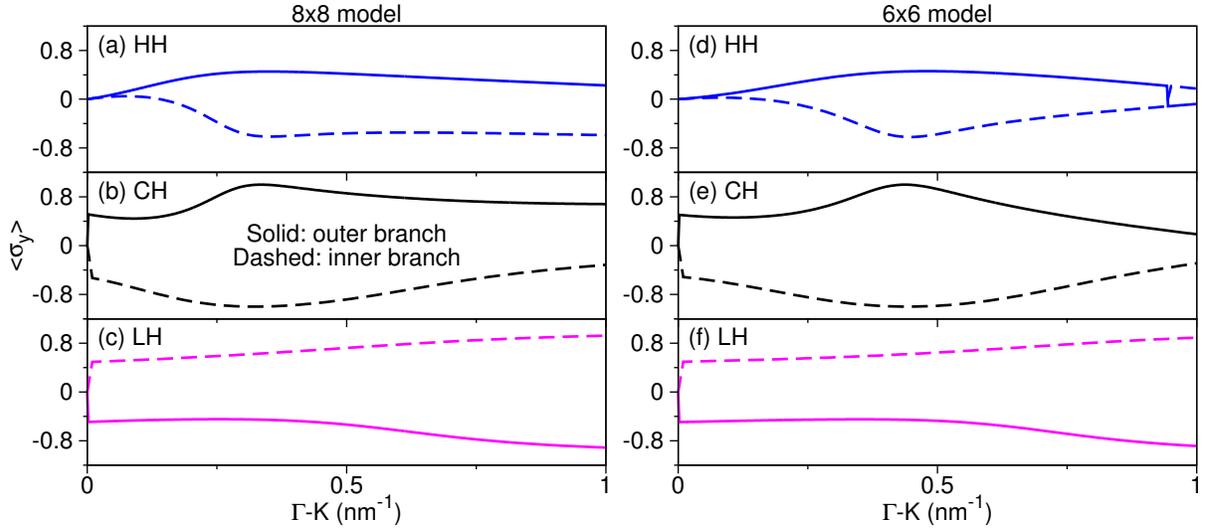}
\caption{(Color online) Spin expectation value in $y$ direction, $\left\langle\sigma_y\right\rangle$, 
along $\Gamma$-K for the (a-c) 8$\times$8 and (d-f) 6$\times$6 model for HH, CH 
and LH bands. The values of $\left\langle\sigma_x\right\rangle$ and $\left\langle\sigma_z\right\rangle$ 
are zero along $\Gamma$-K. Solid (dashed) lines indicate the outer (inner) branch 
of the bands, as presented in Fig. 2(a) and 2(b) of the main paper.}
\label{fig:InAs_6x6_8x8_st_GK}
\end{center}
\end{figure}

In Fig.~\ref{fig:InAs_6x6_8x8_st_GK} we show the spin expectation value in $y$ 
direction, $\left\langle\sigma_y\right\rangle$, along $\Gamma$-K direction for 
InAs valence band. Despite the differences for the band structure and the spin 
splittings, the spin orientation follows the same trends and signs of the 8$\times$8 
model, except for the crossing in the HH band that happens for a smaller $k$ value.

\subsection*{B. InP}

\begin{figure}[h!]
\begin{center} 
\includegraphics{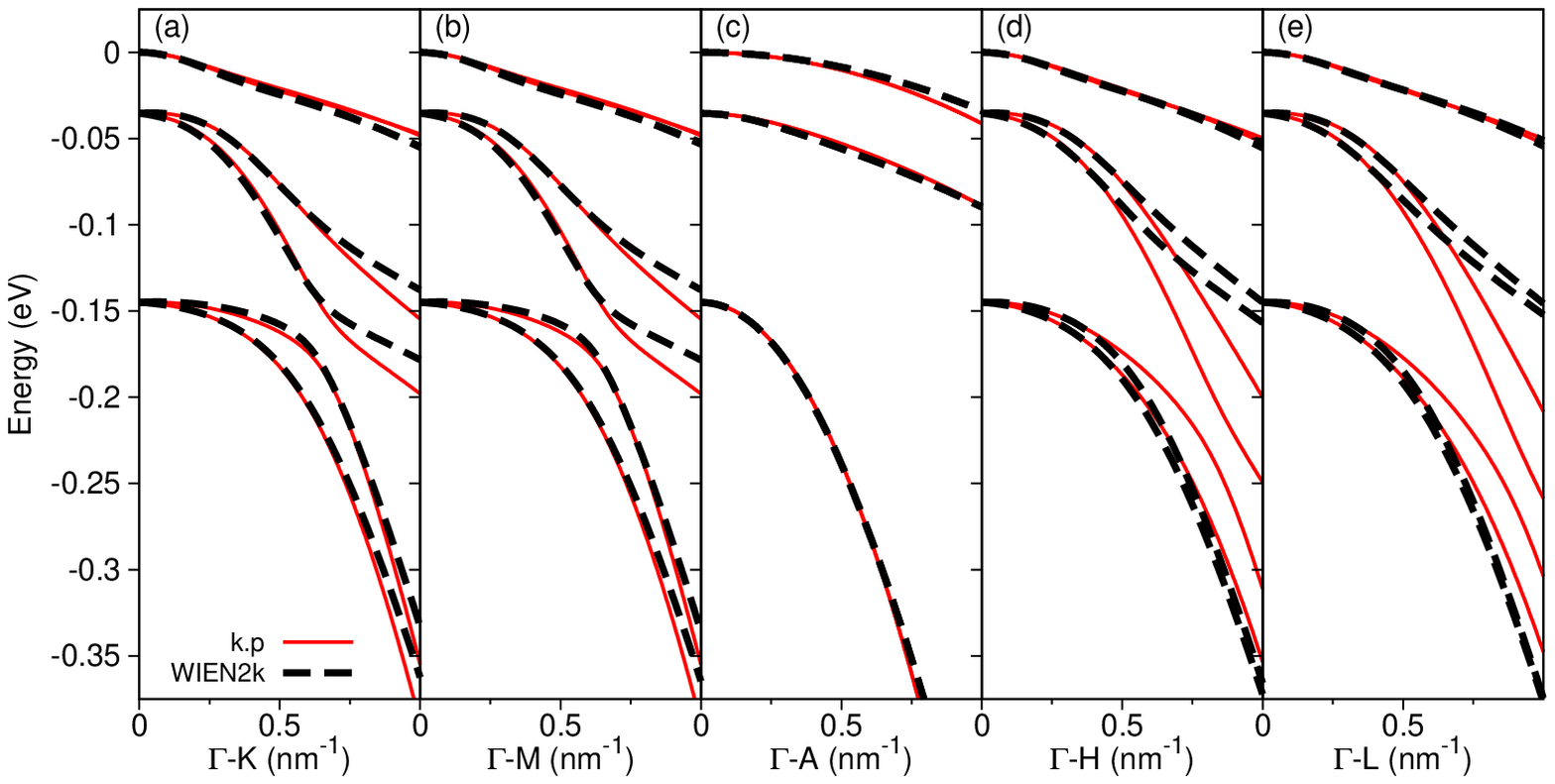}
\caption{(Color online) Same as Fig.~\ref{fig:InAs_6x6_bs} but for InP.}
\label{fig:InP_6x6_bs}
\end{center}
\end{figure}

\begin{figure}[h!]
\begin{center} 
\includegraphics{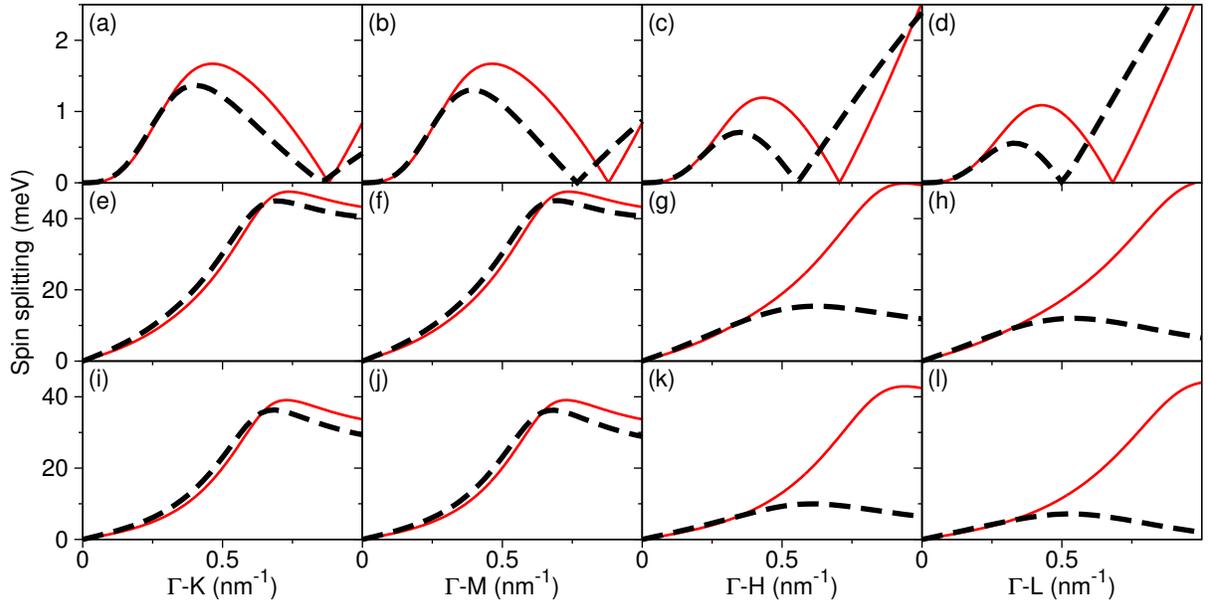}
\caption{(Color online) Same as Fig.~\ref{fig:InAs_6x6_sp} but for InP.}
\label{fig:InP_6x6_sp}
\end{center}
\end{figure}

For InP, the fitting approach for the valence band provides a reasonable agreement 
for the band structure, Fig.~\ref{fig:InP_6x6_bs}, and the spin splitting, Fig.~\ref{fig:InP_6x6_sp}. 
It is important to note that within this 6$\times$6 model, there is only one parameter, 
$A_6$, that couples $k_{x(y)}$ and $k_z$ wave vectors. Because InP band structure 
along $\Gamma$-H(L) direcion is different from $\Gamma$-K(M) direction, only $A_6$ 
is not capable of correcting this anisotropy. Therefore, we see larger differences 
for the band structure and spin splittings along $\Gamma$-H and $\Gamma$-L directions. 

\begin{figure}[h!]
\begin{center} 
\includegraphics{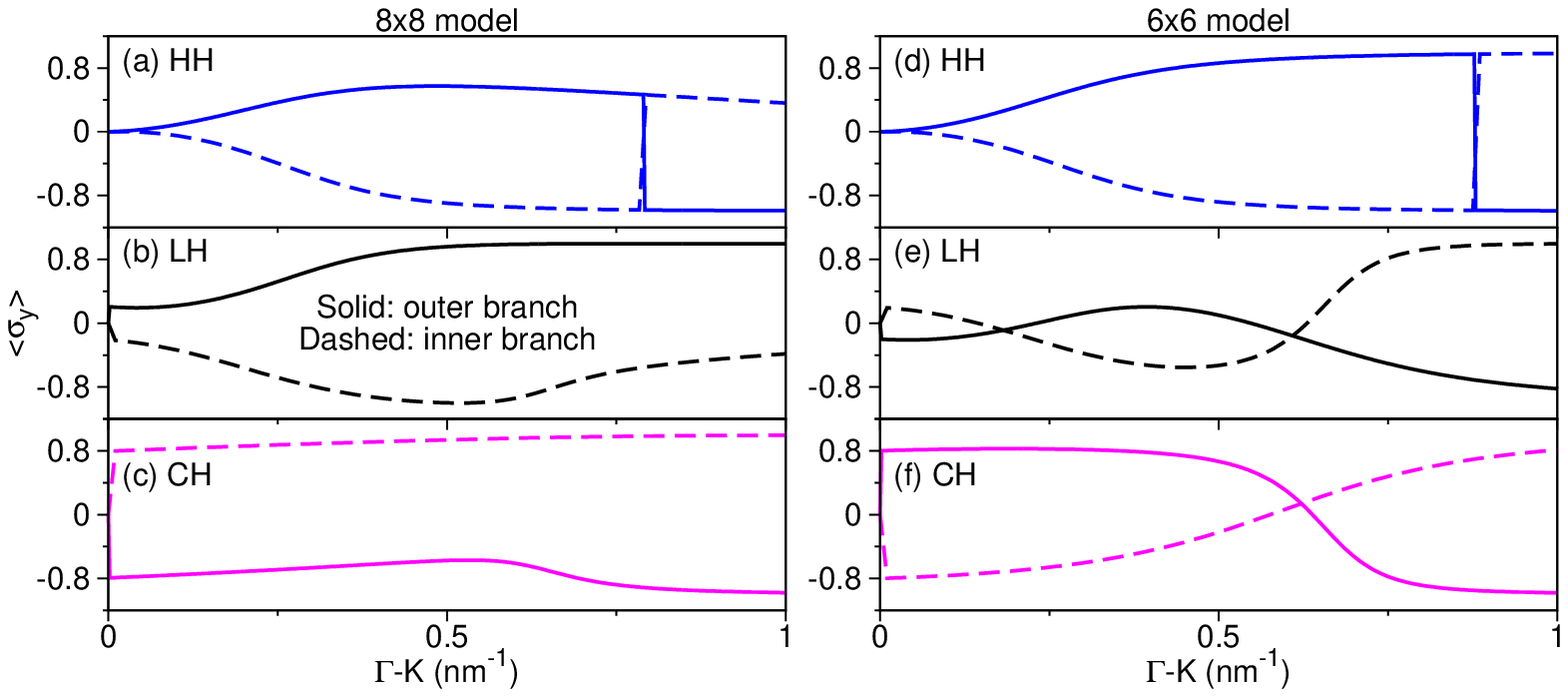}
\caption{(Color online) Same as Fig.~\ref{fig:InAs_6x6_8x8_st_GK} but for InP.}
\label{fig:InP_6x6_8x8_st_GK}
\end{center}
\end{figure}

In Fig.~\ref{fig:InP_6x6_8x8_st_GK} we present $\left\langle\sigma_y\right\rangle$ 
along $\Gamma$-K direction for the valence band of InP. Although HH band in the 6$\times$6 
model shows a similar behavior, LH and CH bands do not follow the trends from the 
8$\times$8 approach. The correct description of InP spin texture, by imposing 
$A_7 > \alpha_1$ in the fitting process (this is seen for InAs parameters), drastically 
compromises the band structure and spin splittings. Therefore, we also suggest the 
use of 8$\times$8 model to treat InP.

%===============================================================================


\begin{thebibliography}{99}
\bibitem{note:doublegroup}
As far as we are concerned, the only paper in the literature that provides 
a double group description of {\bf k.p} model (for cubic crystals) is W. J. Elder, 
R. M. Ward, and J. Zhang, Phys. Rev. B {\bf 83}, 165210 (2011).
\bibitem{note:dirac}
Explicitly, $u_{n,\Gamma}(\vec{r}) = \left\langle \vec{r} | c_n \right\rangle$, 
with $n=1,\ldots,8$ running over the energy bands at $\Gamma$ point we are 
describing.
\end{thebibliography}
\end{document}